\documentclass[%
reprint,
superscriptaddress,
showpacs,preprintnumbers,
amsmath,kantlipsum,amssymb
]{revtex4-1}
\usepackage{hyperref}

\usepackage{graphicx}
\usepackage{dcolumn}
\usepackage{bm}
\usepackage{xcolor,lineno}
\usepackage{soul}
\usepackage{hyperref}

\allowdisplaybreaks
\begin{document}
	\title{Photon antibunching control in a quantum dot and metallic nanoparticle hybrid system with non-Markovian dynamics}%
	
	\author{T. Moradi}%
	\email{ta.moradi@sci.ui.ac.ir}
	\affiliation{Department of Physics, Faculty of Science, University of Isfahan, Hezar Jerib Str., Isfahan 81746-73441, Iran}
	\author{M. Bagheri Harouni}%
	\email{m-bagheri@phys.ui.ac.ir}
	\affiliation{Department of Physics, Faculty of Science, University of Isfahan, Hezar Jerib Str., Isfahan 81746-73441, Iran}%
	\affiliation{Quantum Optics Group, Department of Physics, Faculty of Science, University of Isfahan, Hezar Jerib Str., Isfahan 81746-73441, Iran}%
	\author{M. H. Naderi}%
	\affiliation{Department of Physics, Faculty of Science, University of Isfahan, Hezar Jerib Str., Isfahan 81746-73441, Iran}%
	\affiliation{Quantum Optics Group, Department of Physics, Faculty of Science, University of Isfahan, Hezar Jerib Str., Isfahan 81746-73441, Iran}%
	
\begin{abstract}
	Photon-number statistics of the emitted photons from a quantum dot placed in the vicinity of
	a metallic nanoparticle (with either shell or solid-sphere geometry) in the non-Markovian regime is investigated theoretically. In the model
	scheme, the quantum dot is considered as a InAs three-level system in $\Lambda$-type configuration with two
	transition channels. One of channels is driven by
	a polarized classical field while the two channels are coupled to the plasmon modes. Plasmon resonance modes of a nanoshell, in contrast of a nanosphere, are tunable
	at demand frequency by controlling the thickness and the materials of the core and the embedding
	media. The results reveal that the emitted photons from the hybrid system under consideration
	are antibunched. Moreover, the antibunching behavior of the emitted photons can be controlled
	by the geometrical parameters of the system, namely, the quantum dot-metal nano particle separation distance, as well as the system's physical parameters including the detuning frequency
	of the quantum dot transitions with respect to the surface plasmon modes, and the Rabi frequency
	of the polarized driving field. Additionally, the studied system has the potential to be a highly controllable
	single-photon source.
\end{abstract}

\pacs{42.50.Nn, 42.50.Ar, 78.67.Hc, 73.20.Mf}
	\maketitle
	\section{Introduction}
	\par
Single-photon sources, as a crucial ingredient of quantum information technology, are amongst the most widely investigated quantum systems during the last four decades or so \cite{1,2,3,4}. Numerous applications for single-photon sources have been  proposed in the fields of quantum cryptography \cite{6,7,8,9}, quantum repeaters \cite{10, 11}, and quantum computation \cite{12}. In addition, the low-noise nature of nonclassical photons makes them as ideal candidates for application in the fundamental measurement problems \cite{13,14,15}.
	 To date, various theoretical schemes and experimental demonstrations have been carried
out on the generation of single-photon emitters. Some examples include atomic cascade transition in calcium atoms \cite{16}, single ions in traps \cite{17}, parametric down-conversion \cite{19,18}, single molecules coupled to a resonant cavity\cite{20}, semiconductor quantum dots (QDs) \cite{4,21,22,22a}, defect centers in diamond \cite{23,24}, single-walled carbon nanotubes \cite{25}, and photonic crystals \cite{25-1}.

\par Another way of realizing an on-demand single photon source is to use the hybrid systems composed of a semiconductor QD and a metal nanoparticle (MNP) \cite{26,sergy}. It is well known that the environment of an emitter changes its decay rate (Purcell effect) \cite{26-1}. The interaction of an MNP with light leads to non-propagating excitations of the conduction electrons of the MNP; the quanta of these excitations are called localized surface plasmons (LSPs) \cite{26-2}. The evanescent near-field associated to the LSPs increases the local density of states (LDOS) around the MNP dramatically \cite{27,28,29,30,30-1,30-2}. When a QD, as an emitter, is located in the evanescent field of an MNP, the decay rate of the QD is affected through the LDOS and increases significantly \cite{27,28,29,30-2}. The physical features of the QD-MNP system can be controlled by the geometry of the hybrid system, i.e., the shape of the MNP and the QD-MNP separation distance. Depending on the geometry, the system can operate either in the strong-coupling or the weak-coupling regime and exhibits the non-Markovian \cite{27,31} or Markovian behavior \cite{30}.
     \par In recent years, significant theoretical studies have been performed on the photon-number statistics of emitted light from a variety of hybrid systems in order to provide efficient methods for the controllable generation of antibunched photons. In Ref. \cite{31-1} the conditions for the realization of photon antibunching of molecular fluorescence in a hybrid system of a single molecule and a plasmonic nanostructure composed of four nanostrips have been theoretically investigated by using the Green's tensor technique \cite{31-2}. The photon-number statistics from the resonance fluorescence of a two-level atom near a metallic nanosphere driven by a laser field with finite bandwidth has also been studied, and it has been shown that the statistics can be controlled by the location of the atom around the metal nanosphere, the intensity and the bandwidth of the driving laser, and detuning from the atomic resonance \cite{32}. In addition, a theoretical framework based on the combination of the field-susceptibility/Green's-tensor technique with the optical Bloch equations has been developed \cite{33} to describe the photon statistics of a quantum system coupled with complex dielectric or metallic nanostructures. By using an approach based on the Green's function method and a time-convolutionless master equation, the dynamics of the photon-photon correlation function in a hybrid system composed of a solid-state qubit placed near an infinite planar surface of a dissipative metal has been studied \cite{27} under the Markov approximation. For a hybrid structure consisting of an optically driven two-level QD coupled to a metallic nanoparticle cluster it has been shown \cite{34} that the single-photon emission can be efficiently controlled by the geometrical parameters of the system.

\par In this paper, we theoretically investigate the photon-number statistics of the light emitted from a single semiconductor QD with $\Lambda$-type configuration in the vicinity of either a metal nanoshell or a solid nanosphere in the non-Markovian regime. The present study follows two main purposes. First, we aim to introduce the hybrid system as a nonclassical photon source. Second, we intend to investigate whether the photon-number statistics can be controlled by the geometrical and physical parameters of the hybrid system.	
Our theoretical description of the system involves the quantization of electromagnetic field in the presence of an MNP within the framework of the classical dyadic Green's function approach including quantum noise sources which is appropriate for dispersive and absorbing media.

\par  The ohmic nature of metals has a substantial impact on the reduction of the effective QD-MNP interaction, specially at high frequencies due to the interband transitions. One effective way to deal with this phenomenon is to use geometries in which the surface plasmons are induced in low energies. The plasmon resonance frequencies of a nanoshell are adjustable by the thickness of the nanoshell and the dielectric permittivities of the core and the shell materials. Therefore, the resonance frequency modes of a nanoshell can appear at much lower energies than those of a nanosphere \cite{35,48,49,50,51,52}.

\par

The paper is structured as follows. In Sec.II, we first describe the theoretical model of the hybrid system composed of a single QD coupled to an MNP within the framework of the master equation approach. Then, we derive an expression for the normalized second-order autocorrelation function for the photons emitted by the QD. Numerical results and discussions are presented in Sec.III, where we explore the controllability of the photon-number statistics through the physical as well as geometrical parameters of the hybrid system. Finally, we present our conclusions in Sec. V.
		\section{DESCRIPTION OF THE HYBRID SYSTEM }
\begin{figure}\label{f1}
			\includegraphics[width=\linewidth]{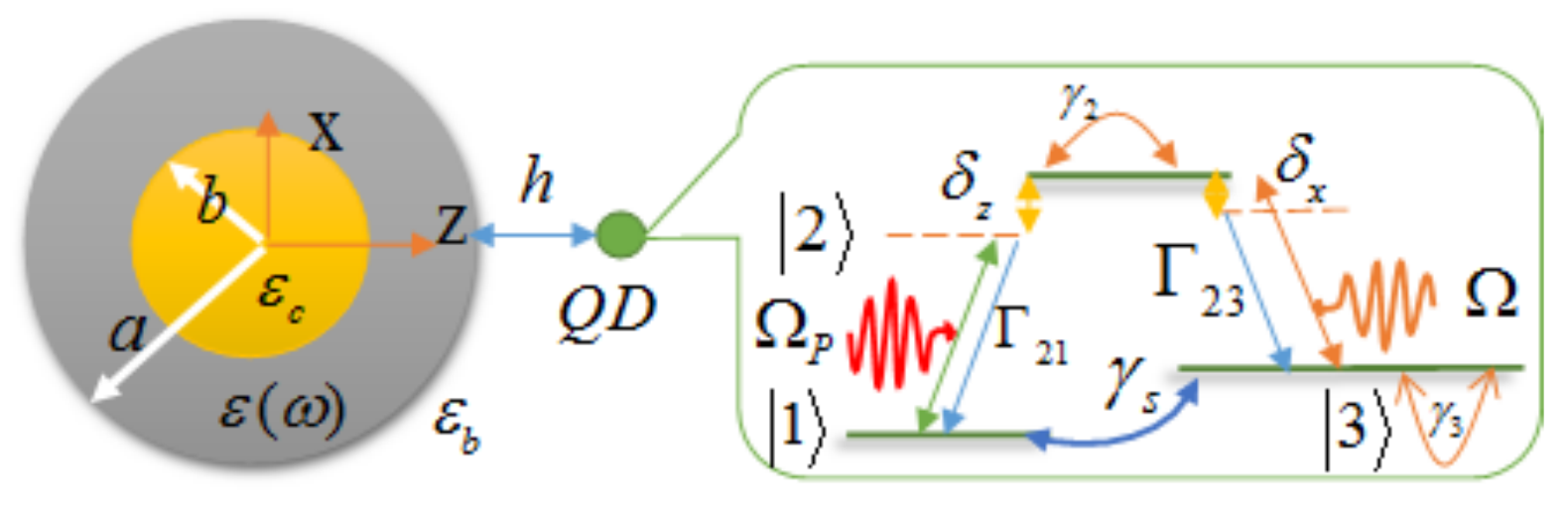}
			\caption{Schematic illustration of the QD-MNP hybrid system under consideration. A QD, as the quantum emitter, is located at distance $h$ away of a nanosphere with radius $R$. The relative permittivity of the metallic nanoshell is frequency-dependent, $\epsilon(\omega)$. The whole system, QD-nanoshell, is embedded in a medium with relative permittivity $\epsilon_{b}$. Here, $\delta_{x}$ and $\delta_{z}$ are detuning of the transitions $| 2 \rangle \longleftrightarrow| 3 \rangle $ and $| 2 \rangle \longleftrightarrow| 1 \rangle $ from the plasmon resonances, respectively. Also, $\Gamma_{23(21)}$ are the decay rates of the state $|2\rangle$ to states $|3(1)\rangle$, $\gamma_{2(3)}$ are the dephasing rates of the states $|2(3)\rangle$, and $\gamma_{s}$ is the spin relaxation of lower energy states. }
			\label{fig:1}
		\end{figure}

		As shown in Fig.\ref{fig:1}, the physical system consists of a QD which is located at distance $h$ from the surface of a nanoshell of inner radius $b$, outer radius $a$, and frequency-dependent permittivity $\epsilon (\omega )$. The QD-nanoshell system is embedded in a homogeneous background medium with relative permittivity $\epsilon_{b}$. As a realistic example like references \cite{27-1,27-2,27-3}, we choose a $\Lambda $-type three-level InAs QD with permittivity $\epsilon_{QD}=11.58$ \cite{55,54}. The discrete energy states  $| 2 \rangle $ and $| 3 \rangle $ have energies $\hbar {{\omega }_{2}}$ and $\hbar {{\omega }_{3}}$, respectively, with respect to the state $| 1 \rangle $, such that $\hbar {{\omega }_{2}}>\hbar {{\omega }_{3}}>\hbar {{\omega }_{1}}$.
	\par 	It has been shown experimentally \cite{54} that the dipole moments $d_{21}$ and $d_{23}$ in the InAs QD are aligned perpendicular to each other. Thus, the atomic dipole moment operator is taken as $\hat d=d(|2 \rangle\langle 1| \hat z+|2 \rangle\langle 3| \hat x )+H.c. $ in which $d$ is real. Moreover, we consider the classical driving field as a $x$-polarized field. Therefore, the transition $| 2 \rangle \longleftrightarrow| 3 \rangle $ is coupled to the classical driving field with Rabi frequency $\Omega$, and detuning $\Delta$. On the other hand, it couples to the x component of surface plasmon modes on the MNP with detuning $\delta_{x}$.
		
		The transition $| 2 \rangle \longleftrightarrow| 1 \rangle $ is coupled to the elementary z-excitations of the MNP with frequency $\Omega_{p}$ and detuning $\delta_{z}$. The classical driving field does not couple to the transition $| 2 \rangle \longleftrightarrow| 1 \rangle $ because they are aligned along the perpendicular directions.
		\subsection{System Hamiltonian   }
		The total Hamiltonian of the whole system can be written as
		\begin{linenomath*}\begin{align}\label{eq:1}
		& \hat {H}=\hat {H}_{QD}+\hat {H}_{F}+\hat {H}_{Int}+\hat {H}_{D},
		\end{align}\end{linenomath*}
		where $\hat {H}_{QD}$ denotes the free Hamiltonian of the QD, $\hat {H}_{F}$ describes the Hamiltonian of the medium-assisted quantized electromagnetic field, and $\hat {H}_{Int}$ refers to the QD-MNP interaction Hamiltonian which are given, respectively, by
			\begin{subequations}
			\begin{linenomath*}\begin{align}\label{eq:2}
			&{\hat {H}_{QD}}=\sum\limits_{i=1}^{3}{{}}\hbar {{\omega }_{i}}{{\hat{\sigma }}_{ii}},\\&
			\hat{H}_{F}=\sum_{i=x,z}\int{{d^{3}}\bm{r}\int_{0}^{\infty }} \hbar \omega_{i} {\bm{\hat{f}}^{\dagger }}(\bm{r},\omega_{i} ).\bm{\hat{f}}(\bm{r},\omega_{i} )d\omega_{i}\\&	{\hat {H}_{Int}}=-[{{\hat{\sigma }}_{21}}\int_{0}^{\infty }{d\omega_{x} {\bm{d}_{21}}.\bm{\hat{E}}(\bm{r}_{d},\omega_{x} )+H.c.]}\nonumber \\&
			-[{{\hat{\sigma }}_{23}}\int_{0}^{\infty }{d\omega_{z} {\bm{d}_{23}}.\bm{\hat{E}}(\bm{r}_{d},\omega_{z} )+H.c.]}.		
			\end{align}\end{linenomath*}
		\end{subequations}
		Here, ${\hat{\sigma }_{ij}}$ are the Pauli operators, $\hat{\bm{f}}(\bm r,\omega )$ and ${\bm{\hat{f}}^{\dagger }}(\bm r,\omega )$ denote, respectively, the bosonic annihilation and creation operators for the elementary excitations of the lossy metal nanoparticle satisfying the commutation relation $[\hat{\bm{f}}(\bm r,\omega ),\hat{\bm{f}}^{\dagger}(\bm r',\omega' )]=\delta (\omega-\omega')\delta (\bm r-\bm r')$  \cite{37}, and ${d_{ij}}$ is the transition dipole moment between $i$ and $j$ levels. The interaction Hamiltonian of Eq.(2c) has been written in the rotating-wave approximation. Moreover, $\hat{\textbf{E}}(\textbf{r}_{d},\omega)$ is the electric field operator at the position of the QD and is defined in the following.
		\par The part $\hat {H}_{D}$ in the total Hamiltonian of Eq.\eqref{eq:1} accounts for the coupling of the QD to the external driving field and is given by ${\hat {H}_{D}}=\hbar \Omega {{\hat{\sigma }}_{23}}{{e}^{-i{{\omega }_{L}}t}}+H.c.$ where the effective Rabi frequency is defined as $\Omega =\langle {\bm{{E}}_{pump}}({\bm{r}_{d}},{{\omega }_{L}}) \rangle .{\bm{d}_{23}}/\hbar \epsilon_{eff}$ with $\epsilon_{eff}=(2\epsilon_{b}+\epsilon_{QD})/3\epsilon_{b}$ in which $\epsilon_{QD}$ is the dielectric constant of the QD. In this definition, $\bm{E}_{pump}$ contains both the direct classical monochromatic field of frequency  ${{\omega }_{L}}$ and amplitude $\textbf{E}_{L}(\textbf{r}_{d},\omega_{L})$, and the scattered field from the MNP which would be defined through the classical dyadic Green's function $\bm{G}({\bm{r}_{d}},\bm{r}',{{\omega }_{L}})$ as \cite{29}
		\begin{linenomath*}\begin{align}\label{eq:3}
		& \bm{{E}}_{pump}({\bm{{r}}_{d}},{{\omega }_{L}})=\bm{E}_{L}({\bm{r}_{d}},{{\omega }_{L}})+\int_{{{V}_{MNP}}} \bm{G}({\bm{r}_{d}},\bm{r}',{{\omega }_{L}})\nonumber\\
		&\times ({\epsilon_{MNP}}({{\omega }_{L}})-1)\bm{E}_{L}(\bm{r'},{{\omega }_{L}}){e}^{i{\varphi }'}d^{3}\bm{r}'.
		\end{align}\end{linenomath*}
		Here ${\varphi }'$ is the phase change associated with the scattered laser field.
		\par
		Quantization of the electromagnetic fields in the presence of an absorbing and dispersive medium via dyadic Green's function approach leads to an explicit expression for the electric field operator in the following form \cite{37}
		\begin{linenomath*}\begin{align}\label{eq:4}
		& \bm{\hat{E}}(\bm r,\omega )=i\sqrt{\frac{\hbar }{\pi {{\epsilon }_{0}}}}\int{\frac{{{\omega }^{2}}}{{{c}^{2}}}}\sqrt{{{\epsilon }_{I}}(\bm{r}',\omega )}\bm{G}(\bm{r},\bm{{r}'},\omega )\nonumber\\ &\times \bm{\hat{f}}(\bm{r}',\omega )d^{3}\bm{r}',
		\end{align}\end{linenomath*}
		where ${{\epsilon }_{I}}(\bm r,\omega )$ is the imaginary part of the frequency-dependent dielectric permittivity of absorbing medium.
		Also, in this equation $\bm{G}(\bm{r},\bm{{r}'},\omega )$ is the dyadic Green's function of the system describing the system response at $r$ to a point source at $r'$. The
		Green's function is obtained through two contributions, $G(r,r',\omega )={{G}^{0}}(r,r',\omega )+{{G}^{s}}(r,r',\omega )$ where ${{G}^{0}}(r,r',\omega )$
		is the direct contribution from the radiation sources in free-space solution and ${{G}^{s}}(r,r',\omega )$
		is the reflection contribution coming from the interaction of the dipole with the materials.
		 In a frame rotating at the laser frequency $\omega _{L}^{x}$ the total Hamiltonian of Eq.\eqref{eq:1} reads $\hat H'=\hat H'_{0S}+\hat H'_{0F}+\hat H'_{Int}$ with
		\begin{subequations}
			\begin{linenomath*}\begin{flalign}\label{eq:5}
			& \hat {H}'_{0S}=\hbar((\omega_{2}-\omega_{L}^{x}) \hat{\sigma }_{22}+{\omega }_{3} \hat{\sigma }_{33})+\hbar( \Omega \hat {\sigma }_{23}+H.c.),\\
			&\hat {H}'_{0F}=\sum_{i=x,z}\int{{d^{3}}\bm{r}\int_{0}^{\infty }} \hbar \omega_{i} {\bm{\hat{f}}^{\dagger }}(\bm{r},\omega_{i} ).\bm{\hat{f}}(\bm{r},\omega_{i} )d\omega_{i},	\\
			&\hat {H}'_{Int}=-[{\hat{\sigma }_{21}}{{e}^{i{{\omega }_{L}^{x}}t}}\int_{0}^{\infty }d\omega_{x} {\bm{d}_{21}}.\hat{\bm{E}}(\bm{r}_{d},\omega_{x} )+H.c.]\nonumber\\
			&-[{\hat{\sigma }_{23}}{{e}^{i{{\omega }_{L}^{x}}t}}\int_{0}^{\infty }d\omega_{z} {\bm{d}_{23}}.\hat{\bm{E}}(\bm{r}_{d},\omega_{z} )+H.c.].
			\end{flalign}
		\end{linenomath*}
		\end{subequations}		
		\subsection{Dynamics of the system}
		The time evolution of the whole system which is composed of the QD, the MNP, and the coherent laser field, in the interaction picture is determined through the Liouville equation \cite{38,39}
		\begin{linenomath*}\begin{align}\label{eq:6}
		d\hat{\tilde{\rho}}_{T}(t)/dt=-(i/\hbar)[\hat{\tilde{H}}'_{Int}(t),{\hat{\tilde{\rho}}_{T}}(t)].
		\end{align}\end{linenomath*}
		Here, the Hamiltonian and the total density matrix in the interaction picture are defined as:
		$\hat{\tilde{H}}'_{Int}(t)={{e}^{i{\hat {H}'_{0}}t/\hbar }}{\hat {H}'_{Int}}{{e}^{-i{\hat {H}'_{0}}t/\hbar }}$ and ${\hat{\tilde{\rho}}_{T}}(t)={{e}^{i{\hat {H}'_{0}}t/\hbar }}{\hat{\rho} _{T}}(t){{e}^{-i{\hat {H}'_{0}}t/\hbar }}$. The time evolution of the total density matrix $\hat \rho _{T}(t)$ in the Schr\"{o}dinger picture is given by
		\begin{linenomath*}\begin{align}\label{eq:7}
		\frac{d{\hat{\rho} _{T}}(t)}{dt}=-\frac{i}{\hbar }[{\hat {H}'_{0}}(t),{\hat{\rho} _{T}}(t)]+{{e}^{-\frac{i{\hat {H}'_{0}}t}{\hbar }}}\frac{d{{\hat{\tilde{\rho}}}_{T}}(t)}{dt}{{e}^{\frac{i{\hat {H}'_{0}}t}{\hbar }}}.
		\end{align}\end{linenomath*}
		By formally integrating Eq.(\ref{eq:6}), and inserting such a formal solution for $\hat{\tilde{\rho}} _{T}(t)$ on the right hand side of Eq.\eqref{eq:6} we arrive at
		\begin{linenomath*}\begin{align}\label{eq:8}
		&\frac{d{{\hat{\tilde{\rho} }}_{T}}(t)}{dt}=-\frac{i}{\hbar }[{\hat{\tilde{H}}'_{Int}}(t),{\hat{\tilde{\rho} }_{T}}(0)]\nonumber\\&-\frac{1}{{{\hbar }^{2}}}\int_{0}^{t}{d{t}'[}{\hat{\tilde{H}}'_{Int}}(t),[{\hat{\tilde{H}}'_{Int}}({t}'),{\hat{\tilde{\rho} }_{T}}({t}')]].
		\end{align}\end{linenomath*}
		We assume that the states of the system and the reservoir are initially uncorrelated, so that ${{\hat{\rho} }_{T}}(0)=\hat{\rho} (0)\otimes {\hat{R}_{0}}$ in which $\hat{\rho} (0)$ and ${\hat{R}_{0}}$ stand for the initial density matrix of the system and the reservoir, respectively. We also assume that the reservoir is in thermal equilibrium. After partial tracing over the reservoir degrees of freedom and applying a second-order Born approximation, the master equation of the reduced density matrix of the QD in the interaction picture, $\hat{\tilde{\rho}}(t)$, is obtained as
		\begin{linenomath*}\begin{align}\label{eq:9}
		&\frac{d}{dt}{\hat{\tilde{\rho} }}(t)=\\&-\frac{1}{{{\hbar }^{2}}}T{{r}_{R}}\int_{0}^{t}{d{t}'[}{\hat{\tilde{H}}'_{Int}}(t),[{\hat{\tilde{H}}'_{Int}}({t}'),\hat{\tilde{\rho} }({t}') \otimes {\hat{R}_{0}}]],\nonumber
		\end{align}\end{linenomath*}
		where we have used $Tr_{R}\{\hat{\tilde{H}}'_{Int}(t)\hat{R}_{0}\}=0$ because the system-reservoir interaction has no diagonal elements in the representation in which the Hamiltonian of the thermal reservoir is diagonal. By substituting Eq.\eqref{eq:9} into Eq.\eqref{fig:7} and partial tracing over the reservoir variables one arrives at
		\begin{linenomath*}\begin{align}\label{eq:10}
		&\frac{d \hat{\rho} }{d t}=\frac{-i}{\hbar }[{\hat {H}'_{0S}},\hat{\rho} ]\nonumber\\&
		-(1/\hbar^{2})Tr_R\int_{0}^{t}dt' \Bigg\{\hat {H}'_{Int}\hat{\tilde{H}}'_{Int}(t'-t)\hat{\rho}(t)\hat{R}_{0} \nonumber\\&-\hat {H}'_{Int}\hat{R}_{0}\hat{\rho}(t)\hat{\tilde{H}}'_{Int}(t'-t)-\hat{\tilde{H}}'_{Int}(t'-t)\hat{\rho}(t)\hat{R}_{0}\hat {H}'_{Int}\nonumber\\&+\hat{R}_{0}\hat{\rho}(t)\hat{\tilde{H}}'_{Int}(t'-t)\hat {H}'_{Int}\Bigg\}.
		\end{align}\end{linenomath*}
		\par	Since the plasmonic modes are excited at optical frequencies whose energy scales are much higher than the thermal energy $k_{B}T$, it is reasonable to assume that the system is at zero temperature \cite{29,30}. Therefore, we can use the following correlation functions for the reservoir operators: $Tr_{R}[\hat{\bm{f}}_{i}(\bm r,\omega)\hat{\bm{f}}_{i}^{\dagger}(\bm r',\omega')\hat{R}_{0}]=\delta(\bm r-\bm r')\delta(\omega-\omega')$ and $Tr_{R}[\hat{\bm{f}}_{i}^{\dagger}(\bm r,\omega)\hat{\bm{f}}_{i}(\bm r',\omega')\hat{R}_{0}]=0$. On the other hand,  taking a look at Eq.\eqref{eq:10}, one can recognize that this master equation is a non-Markovian master equation in the time convolution form.
		After calculating the integrand on the right hand side of Eq.\eqref{eq:10} explicitly we arrive at
		\begin{linenomath*}\begin{align}\label{eq:11}
		&\frac{d {\hat{\rho} }}{d t}=\frac{-i}{\hbar }[{\hat {H}'_{0S}},\hat{\rho} ]+\int_{0}^{t}d\tau ‌\Big\{\alpha_{z} (-\tau )[-{\hat{\sigma }}_{21}{{\hat{\tilde{\sigma }}}_{12}}(-\tau )\nonumber\\&\hat{\rho} (\tau ) +{{\hat{\tilde{\sigma }}}_{12}}(-\tau )\hat{\rho} (\tau ){\hat{\sigma }}_{21}]+
		\alpha_{x} (-\tau )[-{\hat{\sigma }}_{23}{{\hat{\tilde{\sigma }}}_{32}}(-\tau )\nonumber\\&\hat{\rho} (\tau ) +{{\hat{\tilde{\sigma }}}_{32}}(-\tau )\hat{\rho} (\tau ){\hat{\sigma }}_{23}]+H.c.
		\Big\}+{{L}_{pure}}+{{L}_{spin \; flip}},
		\end{align}\end{linenomath*}
		where $t-t'=\tau$, ${{\hat{\tilde{\sigma }}}_{12(32)}}(-\tau )={{e}^{-i{\hat {H}'_{0S}}\tau/\hbar }} \hat{\sigma}_{12(32)}(0){{e}^{i{\hat {H}'_{0S}}\tau/\hbar }}$. By this definition $\hat{\tilde{\sigma }}_{12}=\hat{\sigma}_{12}(0)e^{i(\omega_{2}-\omega_{L}^{x})\tau}$, for on-resonance driving (i.e., $\omega_{2}-\omega_{3}=\omega_{L}^{x})$, ${{\hat{\tilde{\sigma }}}_{32}}(-\tau )=\dfrac{\hat{\sigma}_{32}(0)}{2}(1+\cos(2\Omega \tau))+
		\dfrac{\hat{\sigma}_{23}(0)}{2}(1-\cos(2\Omega \tau))-\dfrac{i}{2}\sin(2\Omega \tau)(\sigma_{22}-\sigma_{33})$,
		and $\alpha_{x(z)} (-\tau )=\int_{0}^{\infty }{J(\omega_{x(z)}){{e}^{i(\omega_{x(z)} -{{\omega }_{L}^{x}})\tau }}}d\omega_{x(z)} $ in which the spectral density $J_{x(z)}(\omega_{x(z)})$ is defined as
		\begin{linenomath*}\begin{align}\label{eq:12}
		&J_{x(z)}(\omega_{x(z)})=\nonumber\\&\frac{1}{\pi \hbar {{\epsilon }_{0}}{{c}^{2}}}{{\omega_{x(z)} }^{2}}\bm{d}_{21(23)}.\operatorname{Im}\bm{G}({\bm{r}_{d}},{\bm{r}_{d}},\omega_{x(z)} ).{\bm{d}_{12(32)}}.
		\end{align}\end{linenomath*}
		The last two terms on the right hand side of Eq.(\ref{eq:11}) correspond to the pure dephasing and spin relaxation of lower energy levels of the QD, respectively. Here,
		${L}=\sum_{\mu} 2\hat{L_{\mu}} \hat{\rho}\hat{L}^{\dagger}_{\mu}-\hat{L}^{\dagger}_{\mu}\hat {L}_{\mu} \hat{ \rho}- \hat{ \rho} \hat {L}^{\dagger}_{\mu}\hat {L}_{\mu}$ in which the operators $\hat{L}_{2(3)}=\sqrt{\gamma_{2(3)}}|2(3)\rangle\langle 2(3)|$ indicate the pure dephasing and $\gamma_{2}$ and $\gamma_{3}$ are the pure dephasing rates of the QD states $| 2 \rangle$ and $| 3 \rangle$, respectively; while $\hat{L}_{1(3)}=\sqrt{\gamma_{s}}|1(3)\rangle\langle 3(1)|$ are the operators of spin relaxation and $\gamma_{s}$ indicates the spin relaxation of lower energy states \cite{54-1}. Furthermore, in deriving Eq.(\ref{eq:11}) we have made use of the relation $\int{{{\epsilon }_{I}(\bm{r},\omega)}\frac{{{\omega }^{2}}}{{{c}^{2}}}}\bm{G}({\bm{r}_{d}},\bm{r}',\omega ){\bm{G}^{*}}(\bm{r}',{\bm{r}_{d}},\omega )d^{3}\bm{r}'=\operatorname{Im}\bm{G}({\bm{r}_{d}},{\bm{r}_{d}},\omega )$ \cite{37}.
		\subsection{Photon-number statistics}
		\par One of the main purposes of the present contribution is to explore the photon-number statistics of the light emitted from the QD-MNP hybrid system. In particular, we intend to analyze the influence of the geometry of the MNP on the statistical properties of the emitted photons. The photon-number statistics can be determined by the normalized second-order photon autocorrelation function, i.e., the conditional probability of detecting the second photon at time $\tau$ when the first photon has already been detected at $\tau=0$ \cite{38}. For a given quantum emitter system with an available excited state and a single or multiple channel(s) of relaxation, this autocorrelation function can be written as \cite{33,26-2}
		\begin{linenomath*}\begin{align}\label{eq:13}
		{{g}^{(2)}}(\tau )=\frac{{{{\rho} }_{22}}(\tau )}{{{{\rho} }_{22}}(\infty )},
		\end{align}\end{linenomath*}
		where $\rho_{22}(\tau)=\langle 2 |\hat{\rho}(\tau)|2\rangle$ and $\rho_{22}(\infty)$ indicates the steady-state population of the excited state.
		\par In order to calculate the correlation function $g^{(2)}(\tau)$ for the system under consideration, we apply the Laplace transform method to solve the master equation \eqref{eq:11}. Laplace transforming the both sides of Eq.\eqref{eq:11} yields
		\begin{linenomath*}\begin{align}\label{eq:14}
		&s\hat{\rho} (s)-\hat{\rho} (t=0)=\dfrac{-i}{\hbar}[{\hat {H}'_{0S}},\hat{\rho} (s)]+\nonumber\\&\int_{0}^{\infty}{{e}^{-st}}dt\int_{0}^{t}{d\tau
			\Big\{ \Bigg(	\beta_{z} (-\tau)\Big(-|2\rangle\langle 2|}\hat{\rho} (t-\tau)\nonumber\\&+|1 \rangle\langle 2|\hat{\rho} (t-\tau)|2\rangle\langle 1|\Big)+
		\dfrac{1}{2}\alpha_{x} (-\tau)\Big(-|2\rangle\langle 2|\hat{\rho} (t-\tau)\nonumber\\&+|3 \rangle\langle 2|\hat{\rho} (t-\tau)|2\rangle\langle 3|
		+|2 \rangle\langle 3|\hat{\rho} (t-\tau)|2\rangle\langle 3|\Big)+\nonumber\\&
		\dfrac{1}{4}\eta^{(1)}_{x} (-\tau)\Big(-|2\rangle\langle 2|\hat{\rho} (t-\tau)-|2 \rangle \langle 3| \hat{\rho}(t-\tau)+|3 \rangle\langle 2|\nonumber\\&\hat{\rho} (t-\tau)|2\rangle\langle 3|
		-|2 \rangle\langle 3|\hat{\rho} (t-\tau)|2\rangle\langle 3|-|2 \rangle \langle 2|\hat{\rho} (t-\tau)|2\rangle\langle 3|\nonumber\\&+|3 \rangle \langle 3|\hat{\rho} (t-\tau)|2\rangle\langle 3| \Big)+
		\dfrac{1}{4}\eta^{(2)}_{x} (-\tau)\Big(-|2\rangle\langle 2|\hat{\rho} (t-\tau)+\nonumber\\&|2 \rangle \langle 3| \hat{\rho}(t-\tau)+|3 \rangle\langle 2|\hat{\rho} (t-\tau)|2\rangle\langle 3|
		-|2 \rangle\langle 3|\hat{\rho} (t-\tau)|2\rangle\langle 3|\nonumber\\&+|2 \rangle \langle 2|\hat{\rho} (t-\tau)|2\rangle\langle 3|-|3 \rangle \langle 3|\hat{\rho} (t-\tau)|2\rangle\langle 3| \Big)
		\Bigg)+H.c. \Big\}+
		\nonumber\\&{{\gamma }_{2}}\Big(2| 2 \rangle \langle  2 |\hat{\rho} (s)| 2 \rangle \langle  2 |-| 2 \rangle \langle  2 |\hat{\rho} (s)-\hat{\rho} (s)| 2 \rangle \langle  2 |\Big)+\nonumber\\&{{\gamma }_{3}}\Big(2| 3 \rangle \langle  3 |\hat{\rho} (s)| 3 \rangle \langle  3 |-| 3 \rangle \langle  3 |\hat{\rho} (s)-\hat{\rho} (s)| 3 \rangle \langle  3 |\Big)+\nonumber\\
		& \gamma_{s}\Big( 2| 1 \rangle \langle  3 |\hat{\rho} (s)| 3 \rangle \langle  1 |-| 3 \rangle \langle  3 |\hat{\rho} (s)-\hat{\rho} (s)| 3 \rangle \langle  3 |+\nonumber\\
		&2| 3 \rangle \langle  1 |\hat{\rho} (s)| 1 \rangle \langle  3 |-| 1 \rangle \langle  1 |\hat{\rho} (s)-\hat{\rho} (s)| 1 \rangle \langle  1 | \Big).
		\end{align}\end{linenomath*}
		Here $\beta_{z} (-\tau)={{e}^{i{(\omega_{2}-\omega^{x}_{L})}(\tau)}}\alpha_{z} (-\tau)$, $\eta^{(1)}_{x} (-\tau)={{e}^{2 i \Omega \tau}}\alpha_{x} (-\tau)$, and  $\eta^{(2)}_{x} (-\tau)={{e}^{-2 i \Omega \tau}}\alpha_{x} (-\tau)$. Considering the initial conditions  $\rho_{33}(0)=1,\rho_{22}(0)=\rho_{11}(0)=0$, and $\rho_{11}+\rho_{22}+\rho_{33}=1$ the Laplace transform of the excited-state population is obtained as
	\begin{widetext}
			\begin{eqnarray}\label{15}
			& {{{\rho} }_{22}}(s)=\dfrac{2\operatorname{\mathbb{R}e}\{\dfrac{A(s)P(s)}{N(s)}\}+2\Omega \operatorname{\mathbb{I}m} \{\dfrac{P(s)}{N(s)}\}}{s+\Gamma_{2}(s)-2\operatorname{\mathbb{R}e}\{\dfrac{A(s)M(s)}{N(s)}\}-2\Omega \operatorname{\mathbb{I}m}\{\dfrac{A(s)M(s)}{N(s)}\}}.
			\end{eqnarray}
	\end{widetext}
			For convenience, we define the total spontaneous emission rate of the state $|2\rangle$ as $\Gamma_{2}(s)=\Gamma_{21}(s)+\Gamma_{23}(s)$ in which $\Gamma_{21}(s)=2\operatorname{\mathbb{R}e}\{\beta_{z}(s)\}$ and $\Gamma_{23}(s)=\operatorname{\mathbb{R}e}\{\alpha_{x}(s)+\dfrac{1}{2}(\eta^{(1)}_{x}(s)+\eta^{(2)}_{x}(s))\}$ with
			\begin{subequations}
				\begin{flalign}\label{eq:17}
					&	\beta_{z} (s)=i\int_{0}^{\infty }{\frac{J_{z}(\omega_{z} )}{\omega_{2} -{{\omega }_{z}}+is}}d\omega_{z},\\\label{eq:17-1}
					&	\alpha_{z(x)} (s)=i\int_{0}^{\infty }{\frac{J_{z(x)}(\omega_{z(x)} )}{\omega^{x}_{L} -{{\omega }_{z(x)}}+is}}d\omega_{z},\\\label{eq:17-2}
					&	\eta^{(1)}_{x} (s)=i\int_{0}^{\infty }{\frac{J_{x}(\omega_{x} )}{\omega^{x}_{L} -{{\omega }_{x}}+2 \Omega+is}}d\omega_{x},\\\label{eq:17-3}
					&	\eta^{(2)}_{x} (s)=i\int_{0}^{\infty }{\frac{J_{x}(\omega_{x} )}{\omega^{x}_{L} -{{\omega }_{x}}-2 \Omega+is}}d\omega_{x}.
				\end{flalign}
			\end{subequations}			
			Moreover, the Lamb shift is given by  $\delta\omega_{2}=\operatorname{\mathbb{I}m}\{2\beta_{z}(s)+\alpha_{x}(s)+\dfrac{1}{2}(\eta^{(1)}_{x}(s)+\eta^{(2)}_{x}(s))\}$\cite{41}. In Eq.\eqref{15} the functions $M(s),$ $N(s)$, and $P(s)$ are defined by
				\begin{widetext}
			\begin{subequations}
				\begin{flalign}\label{eq:16}
				& M(s)=(A^{*}(s))^{2}+\Omega^{2}+(\Gamma_{23}(s)-2\gamma_{s})B^{*}(s)‌-(s+4\gamma_{s}+\Gamma_{21}(s))(\dfrac{(A^{*}(s)+i \Omega)^{2}}{(s+4\gamma_{s})}-B^{*}(s)), \\
				&N(s)=(s+\gamma_{2}+\gamma_{3}+\gamma_{s}+\Gamma_{2}(s)-i\delta\omega_{2}(s))(i\Omega+A^{*}(s))+B^{*}(s)(A(s)-i\Omega), \\&P(s)=\dfrac{(A^{*}(s)+i\Omega)^{2}+2\gamma_{s}B^{*}(s)}{s+4\gamma_{s}},
				\end{flalign}
			\end{subequations}
	\end{widetext}
		where $A(s)=\dfrac{1}{4}(\eta_{x}^{(2)}(s)-\eta_{x}^{(1)}(s))$ and $B(s)=\dfrac{1}{2}\alpha_{x}(s)-\dfrac{1}{4}(\eta_{x}^{(2)}(s)+\eta_{x}^{(1)}(s))$. In order to obtain the time evolution of the excited state population, ${\rho}_{22}(t)$, we should use numerical methods (see section III).
		\section{RESULTS AND DISCUSSION}
		In this section, we present and discuss various numerical results and calculations to analyze the quantum statistics of the photons emitted from the hybrid QD-MNP system under consideration (Fig.\ref{fig:1}). Throughout the calculations,
		we use atomic units ( $\hbar=1$, $4\pi\epsilon_{0}=1$, $c=137$, and $e=1$). For the numerical calculations, we consider a three-level self-assembled InAs QD as the emitter with z- and x-oriented dipole moments of $|d_{21}|=|d_{23}|=0.1$ $e$ nm and $\omega_{21}=0.8046$ eV and  $\omega_{23}=0.8036$ eV \cite{54}. The QD is placed at distance $h$ from the outer surface of a silver nanoshell composed of a spherical core of radius $a$ and permittivity  $\epsilon_{c}$ surrounded by a concentric Ag shell of radius $b$ and frequency-dependent permittivity $\epsilon(\omega)$, embedded in a medium with permittivity $\epsilon_{b}$. \\
		An important ingredient for our numerical calculation is the dielectric permittivity of MNP. At low photon frequencies below the interband transitions region, the permittivity function of metals can be well described by the Drude model. For silver the interband effects already start to occur for energies in excess of 1eV and thus the validity of the Drude model breaks down at high frequencies \cite{42,43}. Therefore, in the numerical calculations we will use the measured dielectric data of silver reported in Ref.\cite{43}. The parameters used for silver are characterized by the plasma frequency of the bulk ${{\omega }_{p}}=9.2$ $eV$ and collision rate of the free electrons $\gamma_{bulk} =0.021$ $eV$ \cite{44}.
			\subsection{LDOS of the system}
			 \par Density of states which change locally around the MNP provides us good insights about the interaction of the QD-MNP and the frequencies of surface plasmons which depend strictly on the orientation of the dipole moment of the QD.
			The scaled LDOS of a hybrid system is defined as ${{\rho }_{zz(xx)}}/{{\rho }_{0}}=\operatorname{\mathbb{I}m}[{{G}_{zz(xx)}}({{r}_{d}},{{r}_{d}},\omega )]/{{\rho }_{0}}$ in which ${{\rho }_{0}}=k_{1}/6\pi $ being the free space density of states \cite{30-1}. Here,
			 \begin{linenomath*}\begin{align}\label{eq:18}
			 &\operatorname{\mathbb{I}m}[{{G}_{zz}}({{r}_{d}},{{r}_{d},\omega)]}=\dfrac{k_{1}}{4 \pi} \operatorname{\mathbb{R}e}\sum\limits_{n=1}^{\infty }{(2n+1)}n(n+1)\nonumber\\
			 &\times{B^{11}_{N}(\omega){[\frac{h_{n}^{(1)}({{k}_{1}}r)}{{{k}_{1}}r}]^{2}}},\\
			 & \operatorname{\mathbb{I}m}[{G}_{xx}({{r}_{d}},{{r}_{d}},\omega)]=\dfrac{k_{1}}{4 \pi} \operatorname{\mathbb{R}e}\sum\limits_{n=1}^{\infty}\dfrac{(2n+1)}{2}\Big[
			 B^{11}_{N}(h^{(1)}_{n}(k_{1}r))^{2}\nonumber\\
			 & +B^{11}_{M}\dfrac{1}{(k_{1}r)^{2}}\Big\{
			 \frac{d({{k}_{1}}r{{h}_{n}^{(1)}({{k}_{1}}r)})}{d({{k}_{1}}r)}{|_{r=a}}
			 \Big\}^{2}\Big],
			 \end{align}\end{linenomath*}
			 where $h^{(1)}_{n}$ is the spherical Hankel function of the first kind and the coefficients $B^{11}_{N}(\omega)$ and $B^{11}_{M}(\omega)$ are given by \cite{45}
			 \begin{eqnarray}\label{eq:20}
			 & B_{N}^{11}(\omega )=-\frac{{{k}_{2}}\partial {{\Im }_{11}}({{\Im }_{21}}-{{\hbar }_{21}}\Re _{N}^{11}(\omega ))-{{k}_{1}}{{\Im }_{11}}(\partial {{\Im }_{21}}-\partial {{\hbar }_{21}}\Re _{N}^{11}(\omega ))}{{{k}_{2}}\partial {{\hbar }_{11}}({{\Im }_{21}}-{{\hbar }_{21}}\Re _{N}^{11}(\omega ))-{{k}_{1}}{{\hbar }_{11}}(\partial {{\Im }_{21}}-\partial {{\hbar }_{21}}\Re _{N}^{11}(\omega ))},\nonumber\\
			 \\	& B_{M}^{11}(\omega )=-\frac{{{k}_{2}} {{\Im }_{11}}({\partial{\Im }_{21}}-{\partial{\hbar }_{21}}\Re _{M}^{11}(\omega ))-{{k}_{1}}{\partial{\Im }_{11}}( {{\Im }_{21}}- {{\hbar }_{21}}\Re _{M}^{11}(\omega ))}{{{k}_{2}} {{\hbar }_{11}}(\partial{{\Im }_{21}}-\partial{{\hbar }_{21}}\Re _{M}^{11}(\omega ))-{{k}_{1}}\partial{{\hbar }_{11}}( {{\Im }_{21}}- {{\hbar }_{21}}\Re _{M}^{11}(\omega ))},\nonumber\\
			 \end{eqnarray}
			 with $
			 {{k}_{1}}=\frac{\omega }{c}\sqrt{{{\epsilon }_{b}}}$, $
			 {{k}_{2}}=\frac{\omega }{c}\sqrt{\epsilon (\omega )}$, $
			 {{k}_{3}}=\frac{\omega }{c}\sqrt{{{\epsilon }_{c}}} $,
			 $
			 {{\Im }_{ij}}={{j}_{n}}({{k}_{i}}{{R}_{j}})$, $
			 {{\hbar }_{ij}}=h_{n}^{(1)}({{k}_{i}}{{R}_{j}}),
			 \partial {{\Im }_{ij}}=\frac{1}{\rho }\frac{d[{{j}_{n}}(\rho )]}{d\rho }{{|}_{\rho ={{k}_{i}}{{R}_{j}}}}$,
			 $\partial {{\hbar }_{ij}}=\frac{1}{\rho }\frac{d[\rho h_{n}^{(1)}(\rho )]}{d\rho }{{|}_{\rho ={{k}_{i}}{{R}_{j}}}},$ and
			 \begin{eqnarray}\label{eq:22}
			 &\Re _{N}^{11}(\omega )=\frac{{{k}_{3}}{{\Im }_{32}}\partial {{\Im }_{22}}-{{k}_{2}}{{\Im }_{22}}\partial {{\Im }_{32}}}{{{k}_{3}}{{\Im }_{32}}\partial {{\hbar }_{22}}-{{k}_{2}}{{\hbar }_{22}}\partial {{\Im }_{32}}},\\
			 &	\Re _{M}^{11}(\omega )=\frac{{{k}_{3}}{{\Im }_{22}}\partial {{\Im }_{32}}-{{k}_{2}}{{\Im }_{32}}\partial {{\Im }_{22}}}{{{k}_{3}}{{\hbar }_{22}}\partial {{\Im }_{32}}-{{k}_{2}}{{\Im }_{32}}\partial {{\hbar }_{22}}}.
			 \end{eqnarray}
			 In addition, $j_{n}$ denotes the spherical Bessel function of the first kind.
			\par
\begin{figure}
		 		\includegraphics[width=\linewidth]{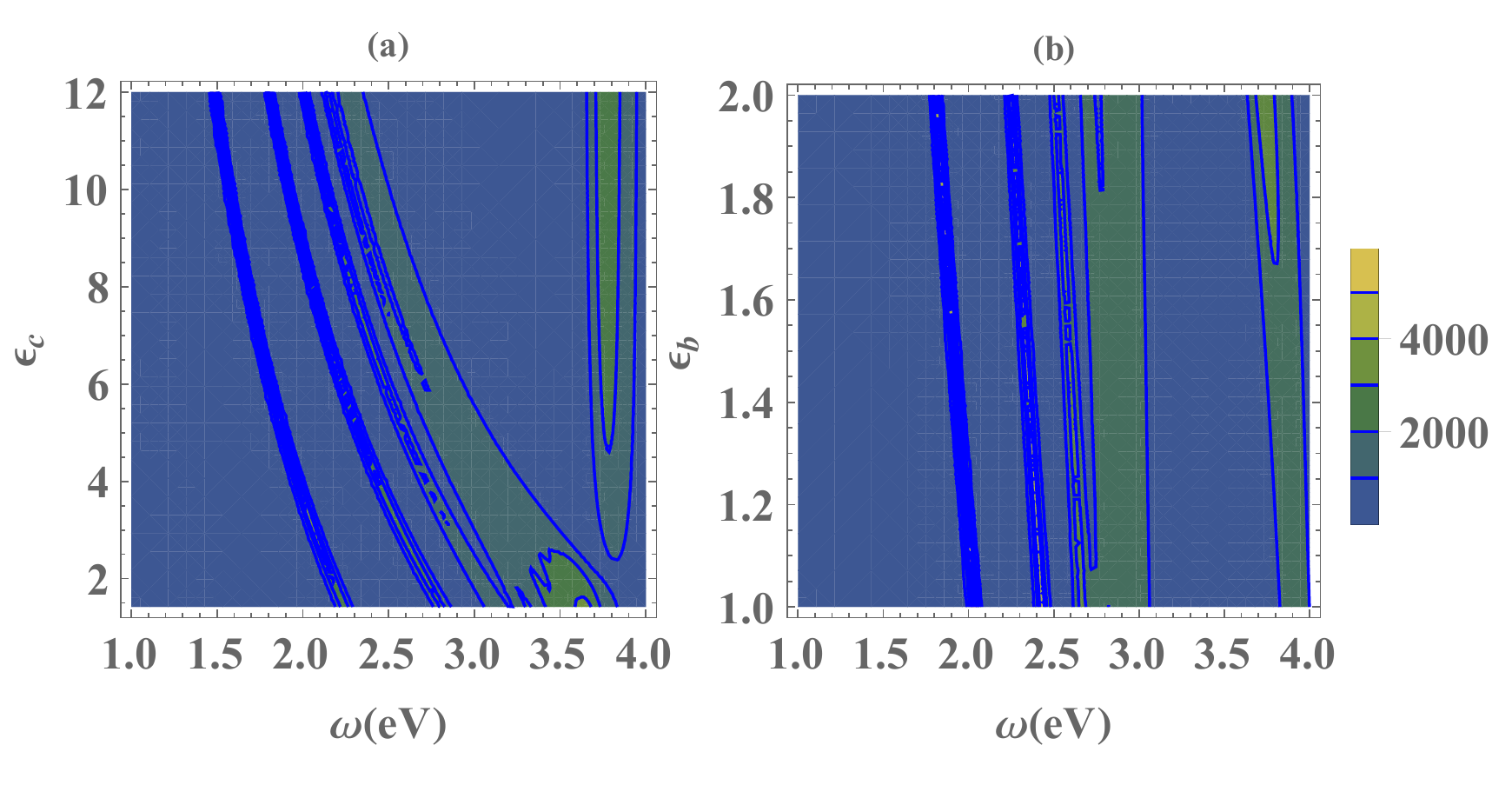}
		 		\caption{Density plot of the scaled LDOS of (a) a (20,16)nm nanoshell versus frequency $\omega$ and the dielectric permittivity of the core $\epsilon_{c}$ when $\epsilon_{b}=1.78$ and (b) a (20,16)nm nanoshell versus frequency $\omega$ and the dielectric permittivity of the background medium $\epsilon_{b}$ when  $\epsilon_{c}=5.4$.}
		 		\label{fig:2}
		 	\end{figure}

 In Fig.\ref{fig:2} we have plotted the scaled LDOS, ${{\rho }_{zz}}/\rho_{0}$, to examine the influence of the core and medium materials on the resonance frequencies of the plasmon modes.
			   As can be seen in Fig.\ref{fig:2}(a), with increasing the core dielectric permittivity $(\epsilon_{c})$ the resonance frequencies of plasmon modes shift toward lower frequencies.
			 Similarly, Fig.\ref{fig:2}(b) shows that the resonance frequencies of the plasmon modes experience a slight redshift once we fix the core dielectric $(\epsilon_{c}=5.4)$ and change the embedding medium dielectric constant from 1 to 2. These results are in agreement with the experimental findings reported in Ref. \cite{49}.
			 \par	  In general, the results show that increasing the dielectric permittivities of the core and the embedding medium lead to the redshift of the plasmon resonances. This behavior has a simple physical interpretation.
			 As is well known, within the quasi-static approach, surface plasmons are collective electromagnetic oscillations at metallic surfaces over a fixed positively charged background, with induced surface charges providing the restoring force. Increasing the dielectric permittivity of the surrounding medium causes the strength of the surface charges to be effectively reduced, leading to a decreased restoring force and consequently, the plasmon energies are lowered. By increasing the core and the medium relative permittivities we adjust the surface plasmon resonance on demand frequency. By tuning the dielectric functions of background and core of the nanoshell to $\epsilon_{b}=\epsilon_{c}=10.95$ \cite{55} the surface plasmons resonance frequencies for $\omega_{n-}$ with $n=1$ is $0.8026$ $eV$ for both $z$ and $x$ surface plasmons which can couple effectively with InAs QD \cite{54}.	
			 \par
\begin{figure}
		 		\includegraphics[width=\linewidth]{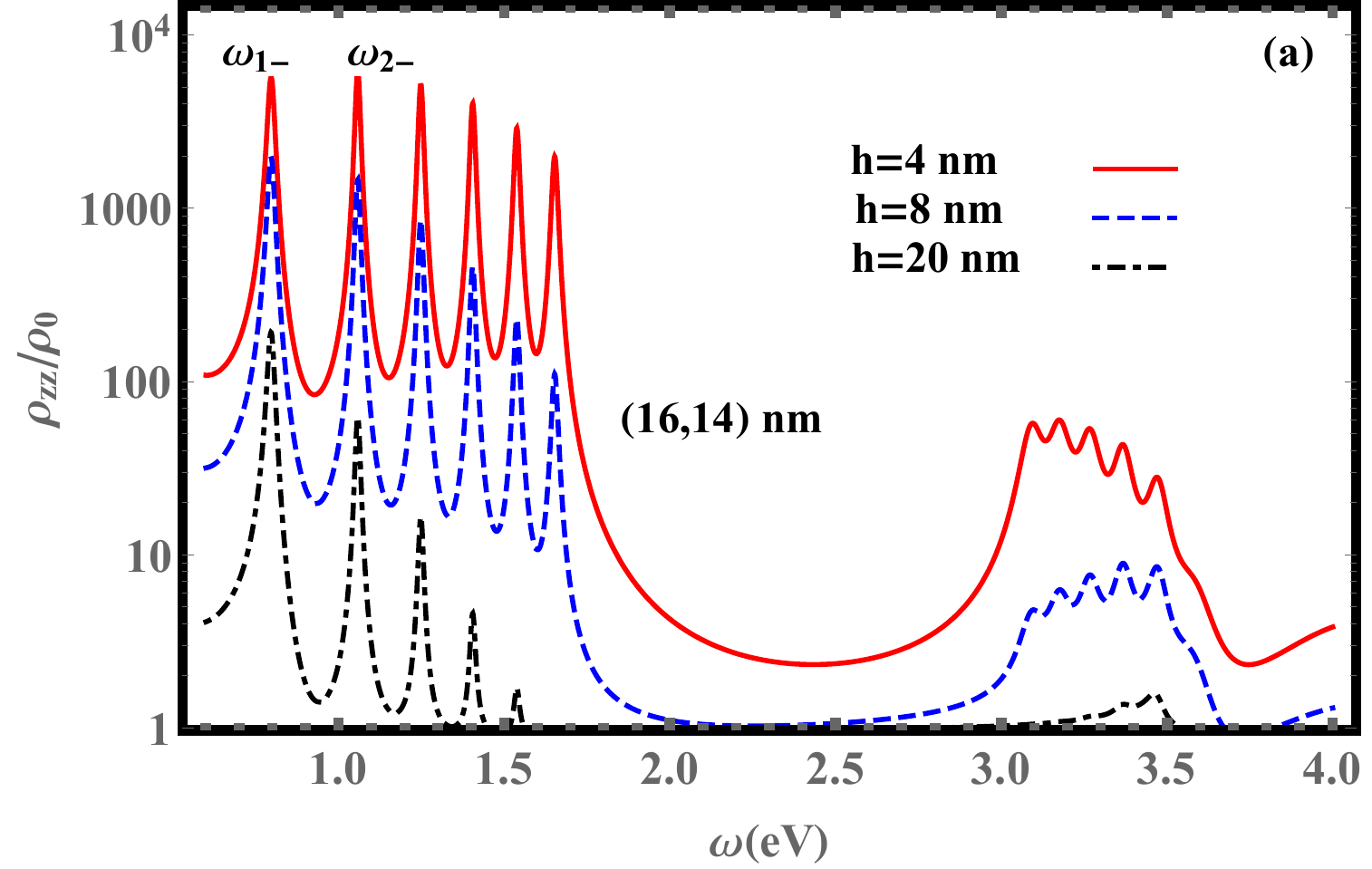}
		 		\includegraphics[width=\linewidth]{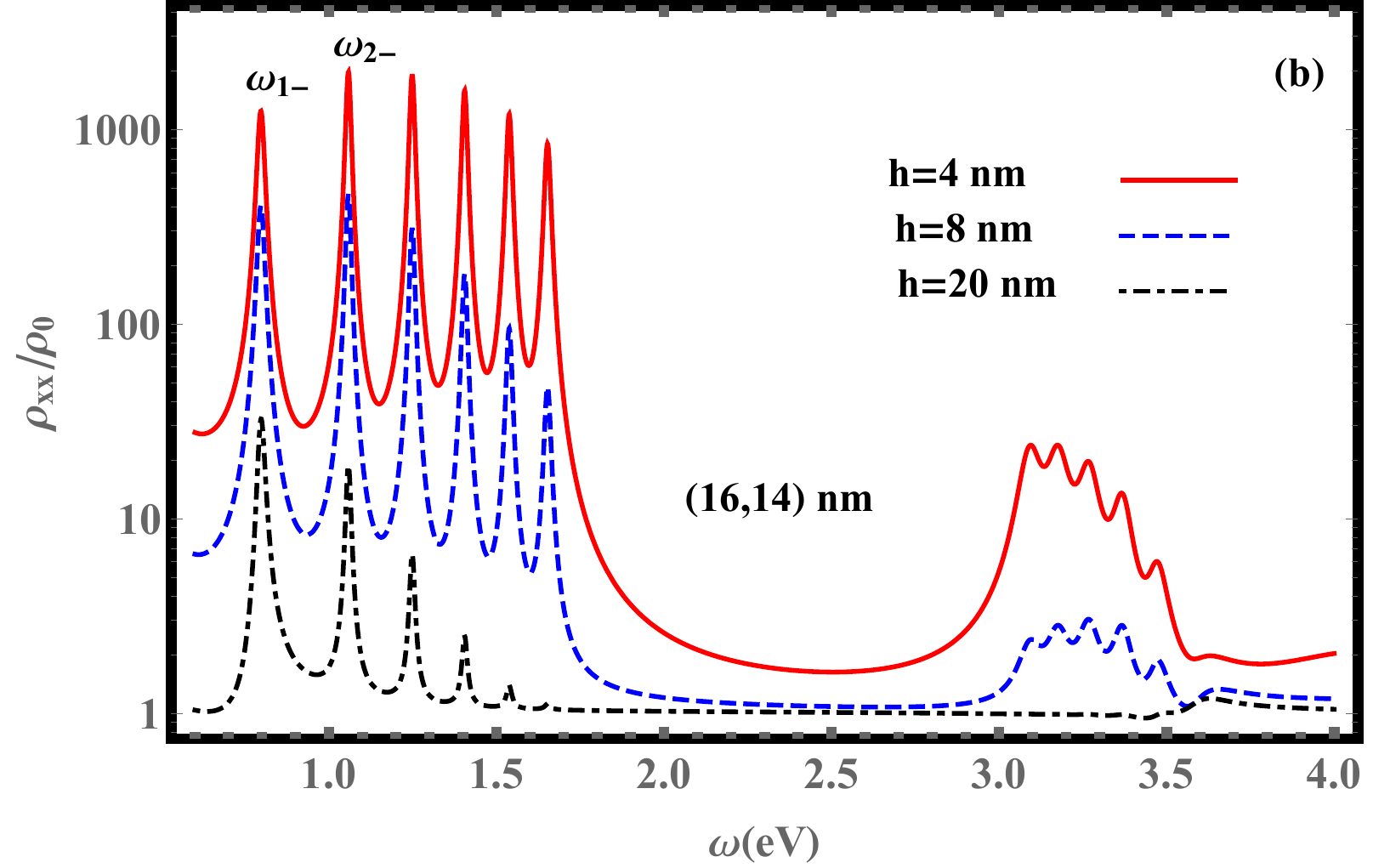}
		 		\caption{Scaled LDOS, (a) ${{\rho }_{zz}}/{{\rho }_{0}}$, (b) ${{\rho }_{xx}}/{{\rho }_{0}}$, versus frequency $\omega$ for a ($16,14$) nm $GaAs/Ag$ nanoshell for different values of $h$. }
		 		\label{fig:3}
		 	\end{figure}

 In Figs.\ref{fig:3}(a) and \ref{fig:3}(b)) we have plotted the scaled LDOS for different QD-MNP separation distances of a $(16,14)$ nm nanoshell composed of a spherical $GaAs$ core of radius $b=14$ nm and permittivity $\epsilon_{c}=10.95$ \cite{47} surrounded by a concentric Ag shell of radius $a=16$ nm and frequency-dependent permittivity $\epsilon(\omega)$, embedded in a medium with permittivity $\epsilon_{b}=10.95$ \cite{47}.
			 \par The resonance frequencies of surface plasmons can be obtained by the poles of the dyadic Green's function. The features of the plasmon resonances in a metallic nanoshell are described by the plasmon hybridization model \cite{46}. In this model, the surface plasmon modes of internal and external surfaces (or core and shell surfaces) are combined in two symmetric and antisymmetric modes. Accordingly, these new branch modes possess low ($\omega^{z(x)}_{n-}$) and high ($\omega^{z(x)}_{n+}$) energies. The thickness of the nanoshell controls the strength of the interaction between the sphere-like and the cavity-like plasmon modes. With increasing the thickness of the nanoshell these modes behave more independently. The important parameters, which possess important influences on the plasmon resonance frequency, are the thickness of MNP $(a-b)$, the core material dielectric constant ($\epsilon_{c}$), and the embedding medium dielectric constant ($\epsilon_{b}$).
		 \par
		According to the hybridization model the plasmon peaks are grouped into two basic branches, i.e., $\Sigma\omega_{n-}$ (plasmon peaks on the left side of Fig.\ref{fig:3}) and $\Sigma\omega_{n+}$ (plasmon peaks on the right side of Fig.\ref{fig:3}). The first left peak belongs to $\omega_{n-}$ with $n=1$ \cite{35}.
		   As can be seen from Figs.\ref{fig:3}(a) and \ref{fig:3}(b) for a fixed aspect ratio $b/a$ the resonance frequencies of plasmon modes are scale-invariant, i.e., although the intensities of plasmon modes get decreased by increasing the QD-MNP separation distance $h$, the energy arrangement remains unchanged for different values of $h$.
		    Thus, by controlling the aspect ratio $b/a$ as well as the materials of the core and medium, one can tune the plasmon resonance frequency on demand.
		    \par Figure \ref{fig:3}(a) shows the enhancement of LDOS due to coupling of the $z$-oriented dipole moment of the QD, i.e., the transition $| 2 \rangle \longleftrightarrow| 1 \rangle $, to the MNP. Similarly, Fig.\ref{fig:3}(b) indicates the coupling of the $x$-oriented dipole moment of the QD, i.e., the transition $| 2 \rangle \longleftrightarrow| 3 \rangle $, to the MNP.
		     Making a comparison between Figs.\ref{fig:3}(a) and \ref{fig:3}(b) reveals that the QD-MNP coupling depends on the orientation of the transition dipole moment of the QD. In the other words, the MNP affects differently on electromagnetic modes with perpendicular polarizations. Thus, the enhancement of LDOS is different for $x$- and $z$-directions and anisotropic Purcell effect happens \cite{30-2}.
		 	\subsection{Spontaneous decay of the QD excited state}
		 	\par
\begin{figure}
		 		\includegraphics[width=\linewidth]{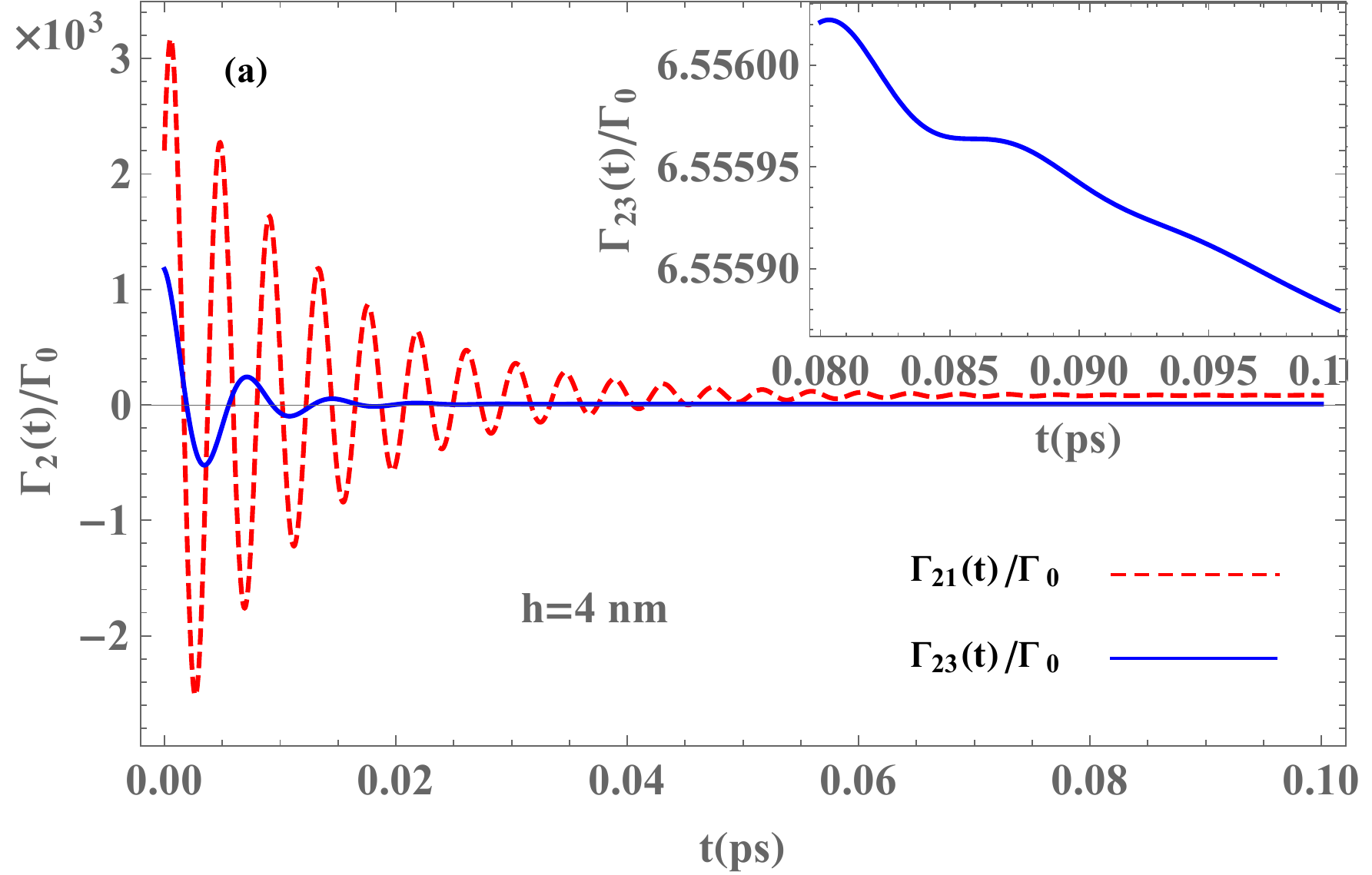}
		 		\includegraphics[width=\linewidth]{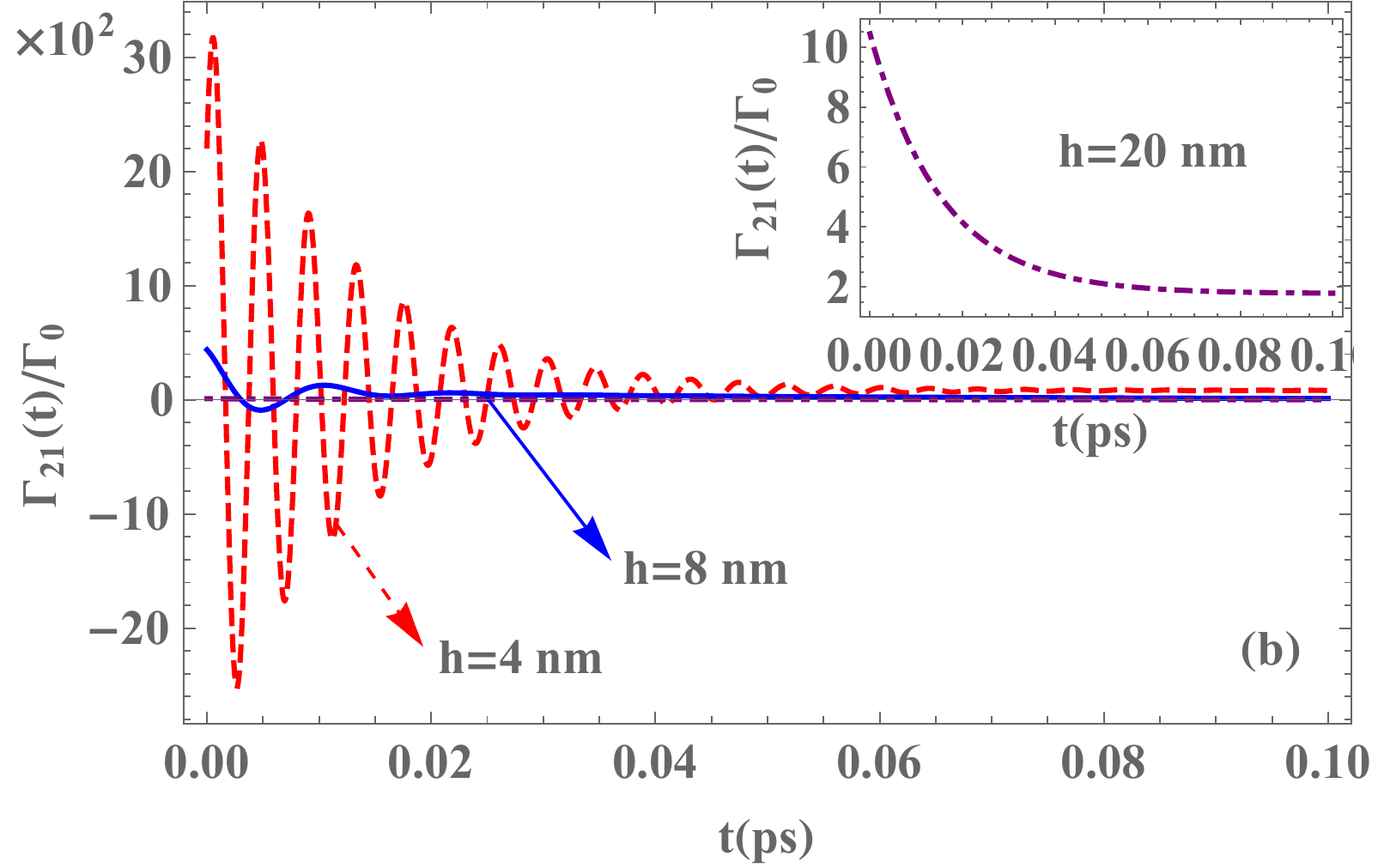}
		 		\caption{Time evolution of the spontaneous decay rate of a QD placed in the vicinity of a (16,14)nm nanoshell. (a) $\Gamma_{21}(t)$ (dashed red curve), $\Gamma_{23}(t)$ (solid blue curve) when the QD is placed at a distance 4 nm from the MNP.		 			
		 			(b) $\Gamma_{21}(t)$ for different values of the QD-MNP separation distance ($h$): $h=4$ nm (dashed red curve), $h=8$ nm (solid blue curve), and $h=20$ nm (dash-dotted purple curve). The inset in (a) represents the long-time relaxation behavior of $\Gamma_{23}(t)/\Gamma_{0}$ and in (b) shows $\Gamma_{21}(t)/\Gamma_{0}$ for $h=20$ nm. Here, the permittivities of the core and the embedding medium are assumed to be equal, $\epsilon_{c}=\epsilon_{b}=10.95$.}
		 		\label{fig:4}
		 	\end{figure}

 It is well known that one of the obvious evidences for the non-Markovian character of relaxation processes such as spontaneous decay of an excited state of a given quantum system is the nonexponential time evolution due to the memory effects. Here, we numerically examine the memory effects on the spontaneous decay of the QD in the system under consideration. For this purpose, we assume that the QD is initially prepared in the excited state $|2\rangle$ and we set $\Omega=0$, i.e., there is an exciton in the QD and no plasmon mode in the MNP. In Fig.\ref{fig:4} we have plotted the scaled spontaneous decay rates of the excited state $|2\rangle$ for both transitions
		 	$| 2 \rangle \rightarrow | 1 \rangle $ and $| 2 \rangle \rightarrow | 3 \rangle $,i.e., 
		 	$\Gamma_{21}(t)/\Gamma_{0}$ and $\Gamma_{23}(t))/\Gamma_{0}$ (determined by the real parts of the inverse Laplace transforms of the parameters $\beta_{z}(s)$, $\eta^{(1)}_{x}(s)$, $\eta^{(2)}_{x}(s)$ and $\alpha_{x}(s)$ given in Eq.\eqref{eq:17}-\eqref{eq:17-3}), versus time $t(ps)$ for different separation distances between the QD and a (16,14)nm nanoshell. 		 	Here, $\Gamma_{0}=(d^{2} \omega_{2}^{3})/(3\pi \epsilon_{0} \hbar c^{3})$ refers to the free-space decay rate of the QD whose typical value is about $10^{6}$ $s^{-1}$ \cite{26-2} . We assume that the detuning between the QD transition $| 2 \rangle \rightarrow | 1 \rangle $ and the surface plasmon mode is near zero ($\delta\simeq 0$) that is the emitter is near resonance with the localized surface plasmon modes and $\omega^{z}_{1-}$=0.8026 eV (see Fig.\ref{fig:3}(a)). 	
		 	The excited QD couples to the MNP and excites a plasmon mode which is nearly on resonance with its energy. The excited resonance plasmon mode re-excites the QD before dissipation to other modes. The oscillatory behavior of $\Gamma(t)$ indicates the re-excitation of the QD after a finite delay time by the reservoir which is the signature of non-Markovian dynamics of the system under study.
			As can be seen from Fig. 4(a) , which corresponds to decay rate of transition $| 2 \rangle \rightarrow | 1 \rangle $ shows completely non-Markovian dynamics of this channel which is directed along the $z$ axis. Moreover, small oscillations of spontaneous decay rate of $|2\rangle\rightarrow|3\rangle$ which is directed along the $x$ axis indicates moderate non-Markovian dynamics of this channel. In Fig.\ref{fig:4}(b) we have plotted the spontaneous decay rate of transition $| 2 \rangle \rightarrow | 1 \rangle $ for $h=4,\;8,$ and $20$ nm. This figure shows that with increasing the QD-MNP separation distance the occupation probability of the QD excited state decays slower in time and its dynamics tends to the Markovian regime for large enough QD-MNP distance. These results are fully consistent with those obtained from Fig.\ref{fig:4}(a).
		 	As can be seen from Figs.\ref{fig:4}(a) and \ref{fig:4}(b),the QD spontaneous decay rates exhibit a damped oscillatory behavior in the course of time evolution and eventually tend to the Markovian decay rate $\Gamma_{0}$ in both channels. It is worth to note that the negative values of decay rates indicate the slow down of the spontaneous decay of the excited state. Physically,
		 	this behavior occurs as the result of the reservoir backaction on the QD which reflects the non-Markovian nature of the dynamics. Furthermore, with increasing the QD-MNP separation distance the amplitude of oscillations decreases such that for $h=20$ nm, the decay rate $\Gamma(t)$ shows no oscillations (the inset plot of Fig.\ref{fig:4}(b)). This is because with increasing the QD-MNP distance, the intensity of the LDOS decreases and consequently, the emitter QD experiences a less structured reservoir.
		  \subsection{Single-photon emission by the three-level QD}
	Here, we are going to investigate the photon-number statistics to better understand the features of the emitted light by a single emitter (QD) in the vicinity of an MNP. The photon-number statistics as well as the quality of a single-photon source are determined by measuring the normalized second-order autocorrelation function of photons, $g^{(2)}(\tau)$, in a Hanbury Brown-Twiss (HBT) setup \cite{53}. Although in the ideal case $g^{(2)}(0)=0$ indicates the antibunching of photons, a value of $g^{(2)}(0)<0.5$ is generally accepted as a criterion for single photon emission \cite{9}.
	\par The second-order autocorrelation function of the photons emitted by the hybrid system under consideration is given by Eq.\eqref{eq:13}. We assume that the QD is initially in its ground state, $| 3 \rangle $, and the x-polarized driving field couples the transition $| 2 \rangle \longleftrightarrow| 3 \rangle $ (along the $x$ axis). Since the transitions $| 2 \rangle \longleftrightarrow| 1 \rangle $ (along the $z$ axis) and $| 2 \rangle \longleftrightarrow| 3 \rangle $ (along the $x$ axis) are in near-resonance with the x- and z-plasmon modes of the MNP, the surface plasmons are excited through the energy of the transitions $| 2 \rangle \longleftrightarrow| 1 \rangle $ and  $| 2 \rangle \longleftrightarrow| 3 \rangle $. In our numerical calculations, we use  the experimental values for the pure dephasing rates $\gamma_{1}$ and $\gamma_{2}$ reported in Ref.\cite{25-1} for InAs QDs at room temperature, i.e.,  $\gamma_{1}=\gamma_{2}=10$ $\mu$eV. The spin transition rate is estimated to be $\gamma_{s}=1\; \mu e V$\cite{27-2}, however, our numerical results show that the spin transition in the system under consideration has a negligible effect on the photon statistics. In what follows, we explore the controllability of the photon-number statistics through the Rabi frequency of the driving field $(\Omega)$, the QD-MNP separation distance ($h$), and detuning frequency of the quantum dot transitions with respect to the surface plasmon modes $(\delta)$.
	 \subsubsection{QD-nanoshell hybrid system}	
	 \par
\begin{figure}
			\includegraphics[width=\linewidth]{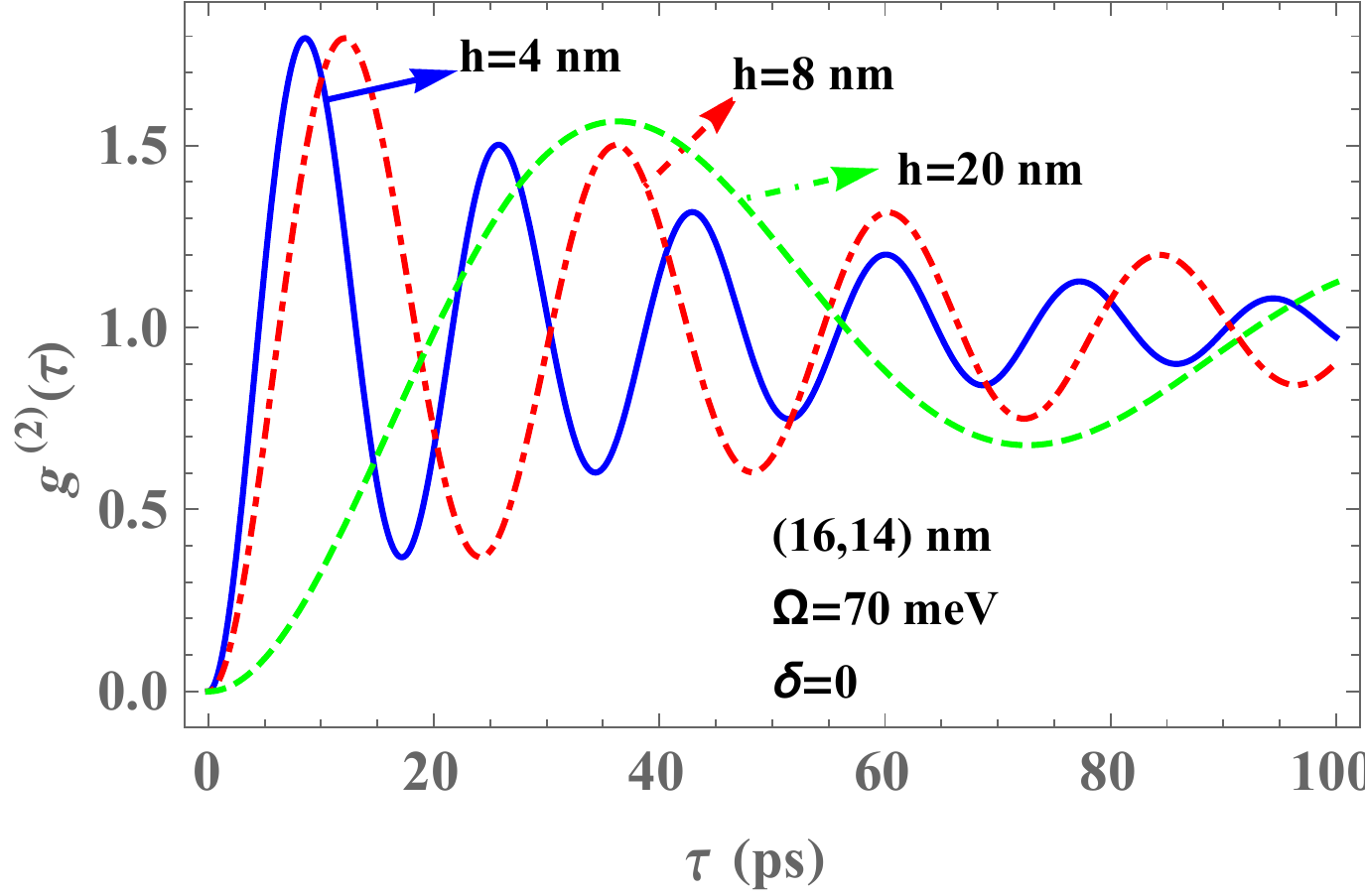}
			\caption{Normalized second-order autocorrelation function, $g^{(2)}(\tau)$, versus the delay time $\tau$ for the photons emitted by a QD located in the vicinity of a (16,14) nm nanoshell for different values of $h$: $h=4$ nm (solid blue curve), $h=8$ nm (dash-dotted red curve), and $h=20$ nm (dashed green curve). Other parameters are the same as those in Fig.\ref{fig:4}.}
			\label{fig:5}
		\end{figure}

	To investigate the effect of the QD-MNP separation distance $h$ on the normalized second-order autocorrelation function of the emitted photons, in Fig.\ref{fig:5} we have plotted $g^{(2)}(\tau)$
	 against the delay time $\tau$ for different values of the QD-MNP separation distance. We assume that the emitter is on resonance with the LSP mode (i.e., $\omega^{z(x)}_{1-}$).
	 This figure demonstrates that by increasing the QD-MNP separation distance the magnitude and the number of oscillations of $g^{(2)}(\tau)$ decrease.
	  This result can be interpreted by the LDOS of the nanoshell. The reduction of the LDOS at frequency $\omega^{z(x)}_{1-}$, (see Fig.\ref{fig:2}(a) and \ref{fig:2}(b)), decreases the coupling strength between the QD and plasmon modes. Consequently, the backaction of the reservoir on the QD and its re-excitation decreases as well. Moreover, by increasing the QD-MNP separation distance the antibunching time increases , because of the enhanced coupling between the QD and the
	   surface plasmon modes compared to the coupling between the QD and the unwated mode. The oscillation periods for $h=4,8,20$ nm are about 19.5, 22, and 78 $ps$, respectively.
	\par
\begin{figure}
		\includegraphics[width=\linewidth]{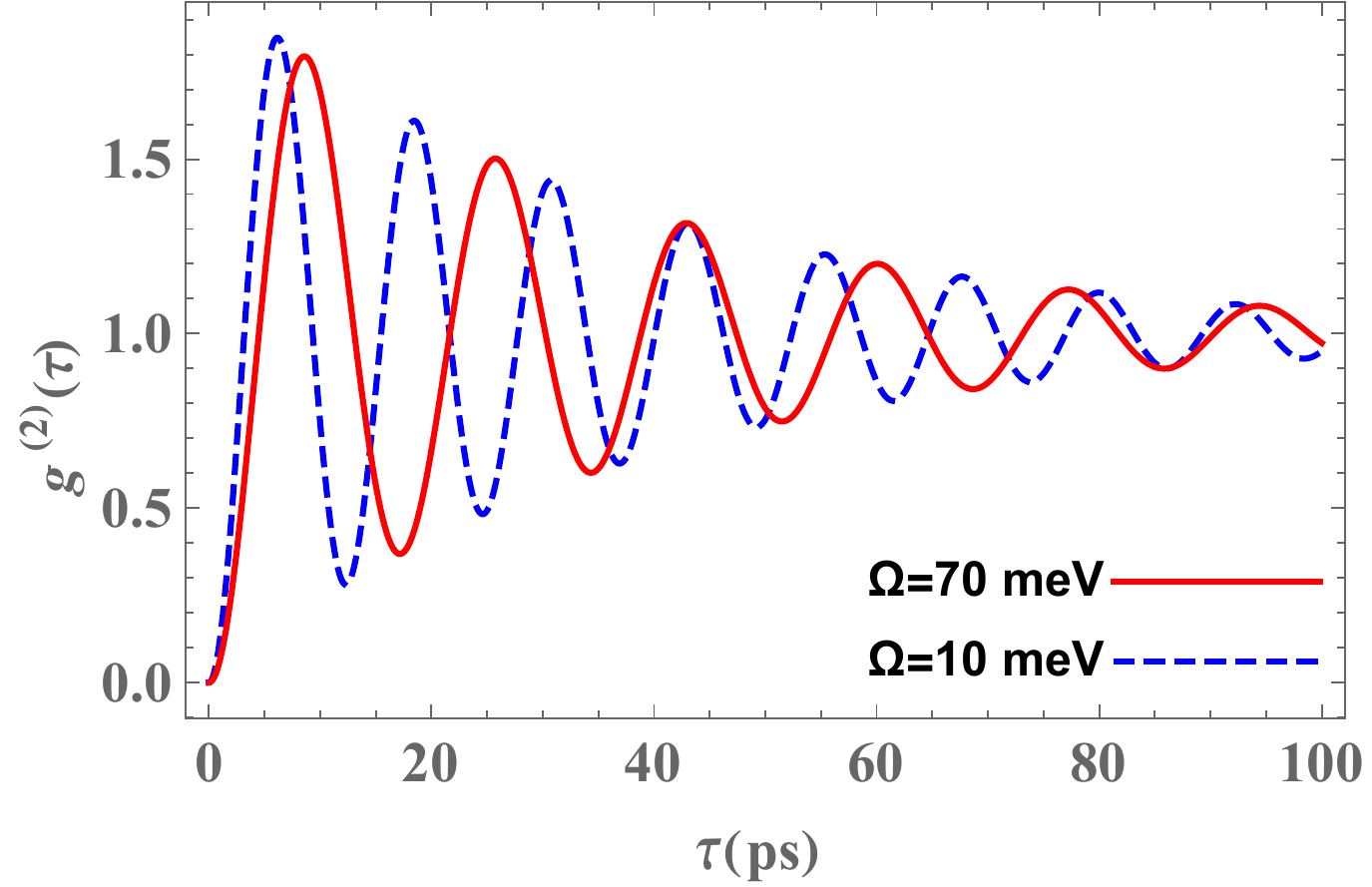}
		\caption{Normalized second-order autocorrelation function, $g^{(2)}(\tau)$, versus the delay time $\tau$ for the photons emitted by a QD located at $h=4$ nm away from a (16,14)nm nanoshell with $\epsilon_{c}=\epsilon_{b}=10.95$, and $\Delta=\omega_{23}-\omega_{L}=0$ for two values of the Rabi frequency of the driving laser: $\Omega=70$ meV (solid red curve),  $\Omega=10$ meV (dashed blue curve). Other parameters are the same as those in Fig. \ref{fig:4}}
		\label{fig:6}
	\end{figure}

  In Fig.\ref{fig:6} we have plotted the autocorrelation function $g^{(2)}(\tau)$ versus the delay time $\tau$ for the different values of the Rabi frequency when the QD is placed at distance $h=4$ nm from a (16,14) nm nanoshell. Furthermore, we have assumed that the driving laser is resonantly coupled to the QD transition $|2\rangle\leftrightarrow|3\rangle$, i.e., $\Delta=0$, and the QD transitions $|2\rangle\leftrightarrow|1\rangle$ and $|2\rangle\leftrightarrow|3\rangle$ are near resonance with the LSP mode of the nanoshell, i.e., $\omega_{1-}=0.8026$ eV; more precisely, $\omega_{21}=\omega^{z}_{1-}+2meV$ and $\omega_{23}=\omega^{x}_{1-}+1meV$ \cite{54}. As can be seen, $g^{(2)}(\tau=0)=0$ and $g^{(2)}(\tau)>g^{(2)}(0)$ as $\tau>0$. This demonstrates the presence of photon antibunching, which is definitely of quantum origin. Moreover, $g^{(2)}(\tau)$  shows an oscillatory dependence on $\tau$. The magnitude of these oscillations decreases as $\tau$ is increased and finally $g^{(2)}(\tau)$ is stabilized at an asymptotic value, $g^{(2)}(\tau)=1$. The oscillatory behavior comes from the non-Markovianity of the system evolution. In addition, the figure shows that increasing the Rabi frequency decreases the coupling strength between the QD and surface plasmon modes and, consequently, the antibunching time increases.
\par
\begin{figure}
		\includegraphics[width=\linewidth]{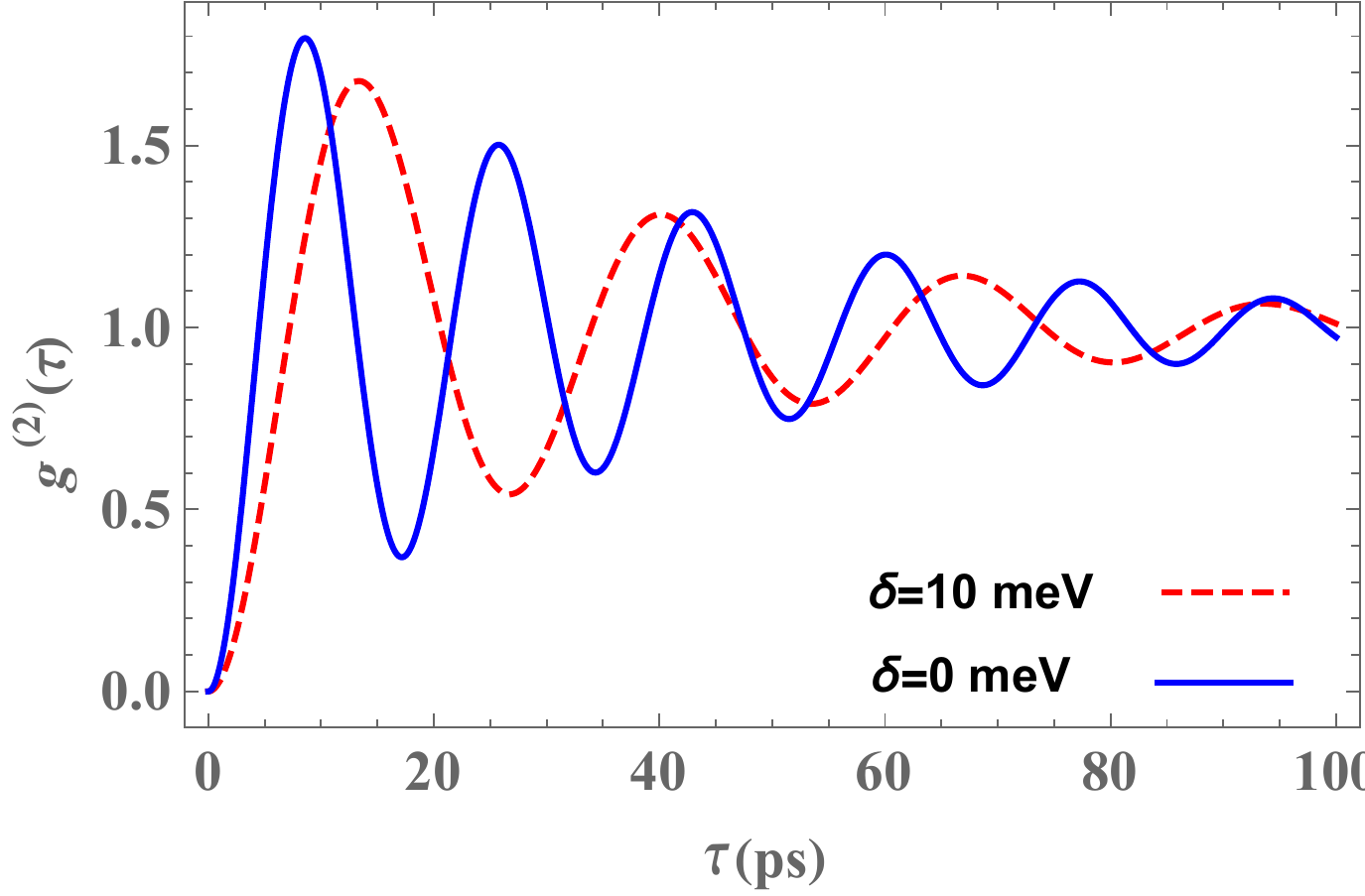}
		\caption{The normalized second-order autocorrelation function, $g^{(2)}(\tau)$, versus the delay time $\tau$ for the photons emitted by a QD located at $h=4$ nm away from a (16,14)nm nanoshell. In this figure solid blue curve corresponds to $\delta=0$ and dashed red curve belongs to $\delta=10$ meV. Other parameters are
			the same as those in Fig.\ref{fig:4}.}
		\label{fig:7}
	\end{figure}

 In the next step, we investigate the impact of detuning frequency between the QD transitions $| 2 \rangle \leftrightarrow | 1 \rangle $ and $| 2 \rangle \leftrightarrow | 3 \rangle $ and two perpendicular surface plasmon modes ($\omega^{z(x)}_{1-}$), i.e., $\delta_{z(x)}$, on the photon-number statistics. In Fig.\ref{fig:7} we have plotted $g^{(2)}(\tau)$ versus the delay time $ \tau$ for two values of the detuning frequency $\delta_{z(x)}$ when the QD is placed at the distance $h=4$ nm from a (16,14)nm nanoshell. One can see from Fig.\ref{fig:7} that with increasing $\delta_{z(x)}$ from zero (solid blue curve) to 0.01 eV (dashed red curve) the antibunching time increases.
 The LDOS of a (16,14)nm nanoshell (Figs.\ref{fig:2}(a) and \ref{fig:2}(b) show that the FWHM of the lowest resonance plasmon mode in energy ($\omega^{z(x)}_{1-}$), which is coupled to the QD transition $|2\rangle \leftrightarrow |1(3)\rangle$, is about $0.1$ eV. By taking some frequency away the summit, the intensity of LDOS is significantly weakened. Consequently, the QD interacts with the unwanted mode and this process leads to the QD dissipation and decreases the coupling strength between QD and surface plasmon modes which makes the Rabi oscillations slower.
\par Based on the results obtained above we can claim that the optimal situation to achieve photon antibunching in a QD-nanoshell hybrid system is one in which $h$=4 nm, the detuning frequency between the QD transition $| 2 \rangle \leftrightarrow | 1(3) \rangle $ and the surface plasmon mode ($\omega^{z(x)}_{1-}$) is zero $\delta=0$, and the Rabi frequency is set to its minimum value. In general, Figs. 5, 6, and 7 show that by decreasing the QD-MNP separation distance and the detuning $\delta$ along with increasing the Rabi frequency $\Omega$ one can achieve optimal photon antibunching in the system under consideration.
	\begin{figure}
		\includegraphics[width=\linewidth]{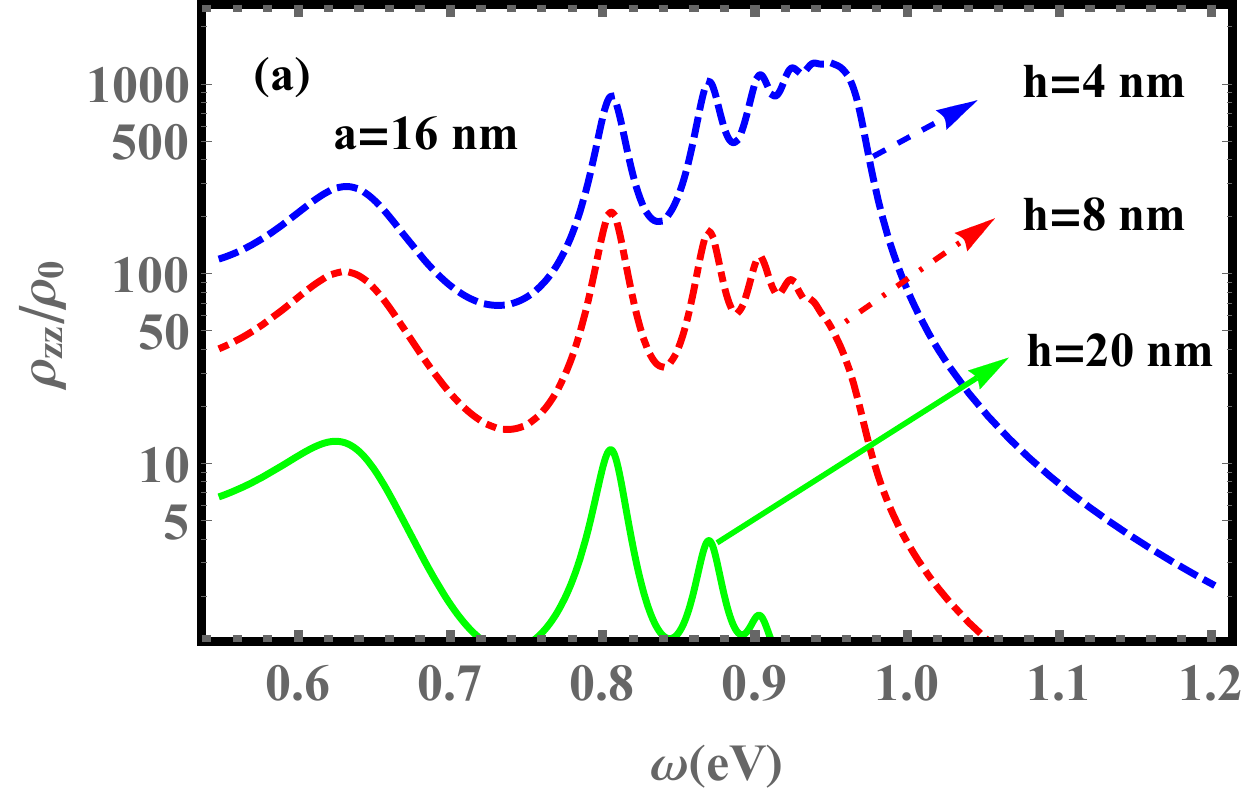}
		\includegraphics[width=\linewidth]{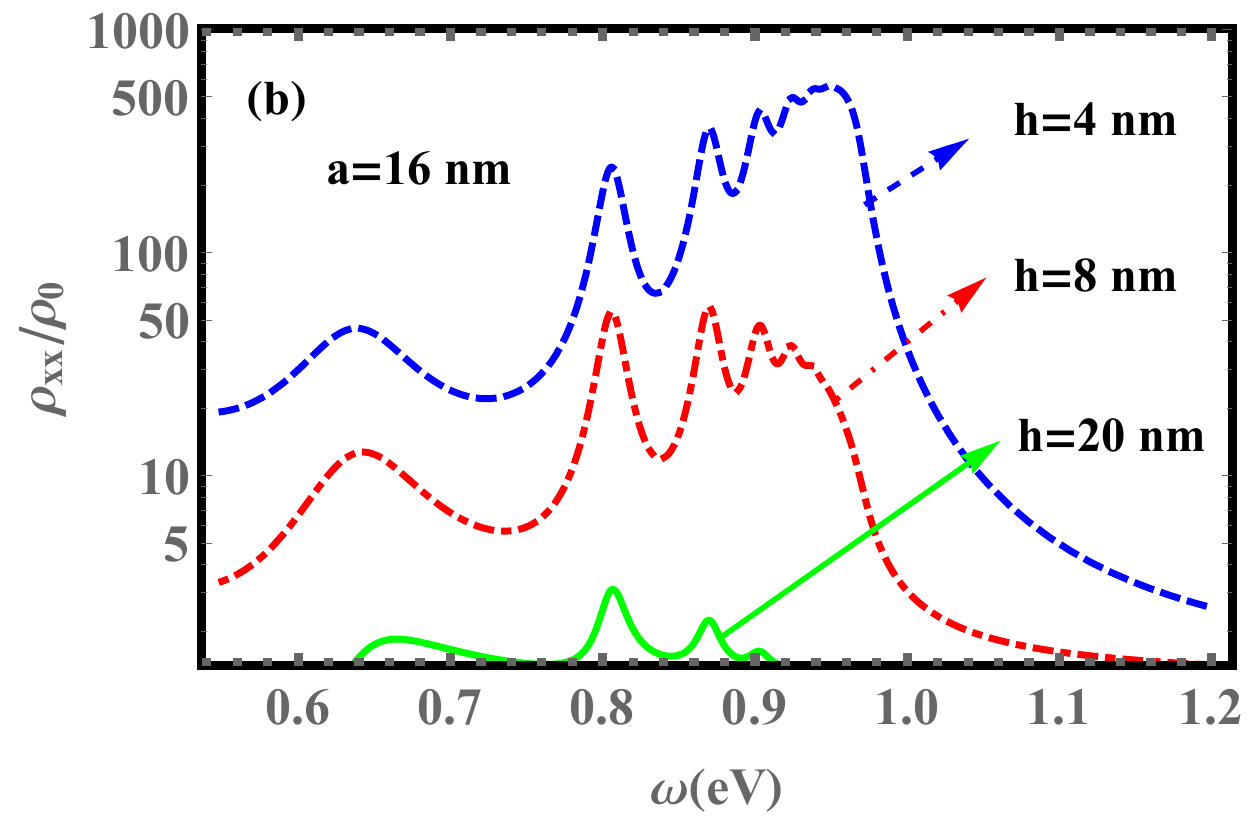}
		\caption{(a) ${{\rho }_{zz}}/{{\rho }_{0}}$ and (b) ${{\rho }_{xx}}/{{\rho }_{0}}$, versus frequency $\omega$ for a $16$ nm $Ag$ nanosphere for different values of $h$: $h=4$ nm (dashed blue curve), $h=8$ nm (dash-dotted red curve), and $h=20$ nm (solid green curve).}
		\label{fig:8}
	\end{figure}	
		  \subsubsection{QD-nanosphere hybrid system}	
		  In the present part, we consider the case in which the InAs QD is placed in the vicinity of a Ag solid-sphere nanoparticle with radius $a$. This case corresponds to the configuration shown in Fig.\ref{fig:1} with $b=0$. We first analyze the dependence of the LDOS of the system on the QD-MNP separation distance, and then investigate the controllability of the photon-number statistics of the light emitted by the system through the QD-MNP separation distance, $h$.
			\par 	In Figs.\ref{fig:8}(a) and \ref{fig:8}(b) we have plotted the scaled LDOS, $\rho_{zz(xx)}/ \rho_{0}$, versus frequency $\omega$ for different values of the QD-MNP separation distance. The whole hybrid system is embedded in a medium with relative permittivity $\epsilon_{b}=78$ which corresponds to a kind of ceramic material, $Ba(Sm,Nd)_{2}Ti_{5}O_{14}$ \cite{56}. The poles of the dyadic Green's function occur at two
			principal modes; the peak at the dipole plasmon mode
			(left peaks in Fig.\ref{fig:8}(a)) is in lower energy
			state compared to the higher-order plasmon modes (right
			peaks in the Fig.\ref{fig:8}(a)). As can be seen, with increasing the QD-MNP separation distance the intensity of dipole plasmon mode increases while the intensities of higher-order plasmon modes decrease. The energy of dipole mode does not change by changing the QD-MNP separation distance; however, the frequency of
			the higher-order
		plasmon modes is slightly blueshifted as the QD-MNP separation distance decreases. The intensity of the LDOS in Figs.\ref{fig:8}(a) and \ref{fig:8}(b) demonstrates that the anisotropic Purcell effect has happened in the nanosphere like in the nanoshell.
			\par
\begin{figure}
		\includegraphics[width=\linewidth]{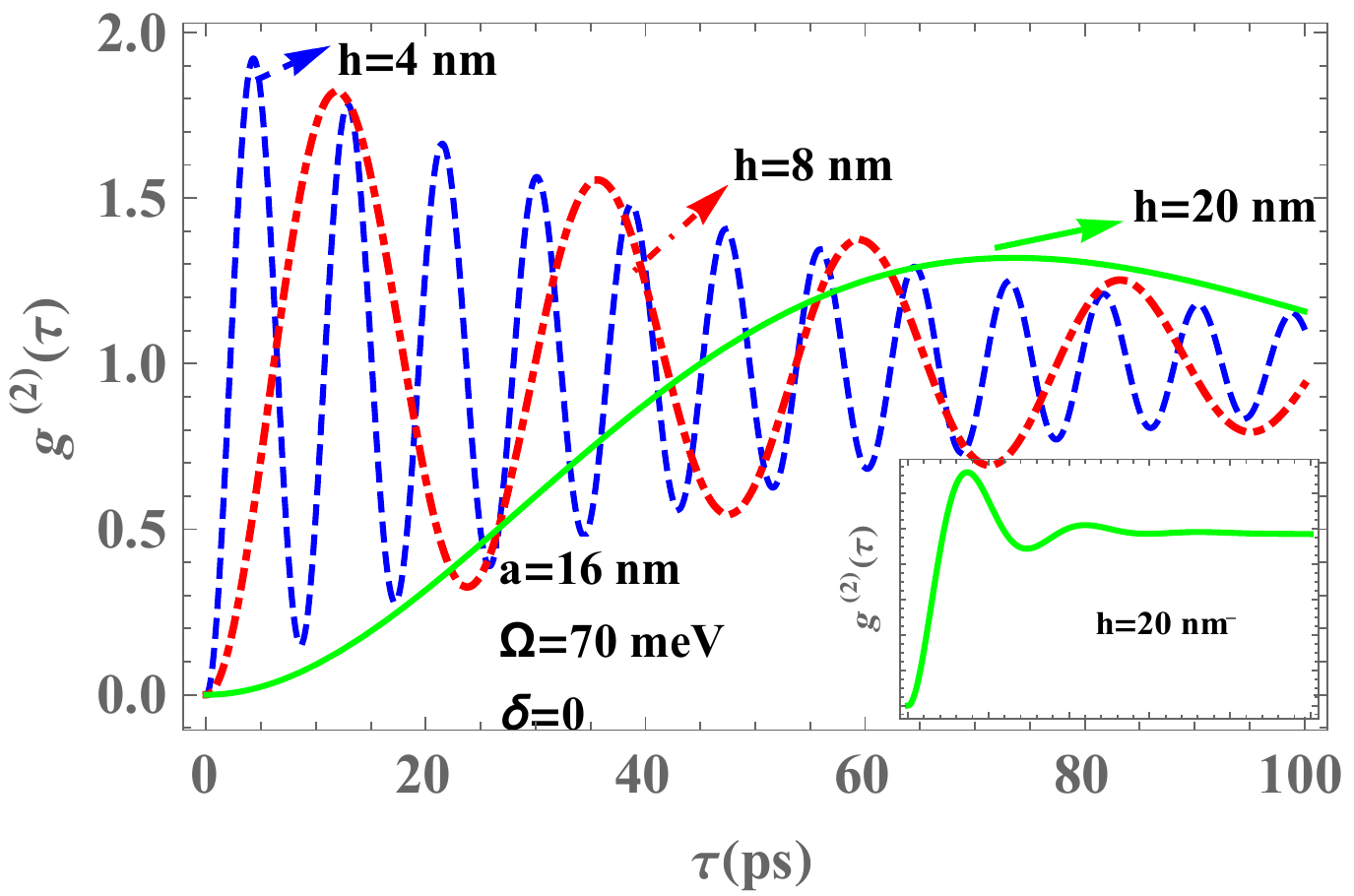}
		\caption{The normalized second-order autocorrelation function, $g^{(2)}(\tau)$,  versus the delay time $\tau$ for different values of the QD-MNP separation distance: $h$: $h=4$ nm (dashed blue curve), $h=8$ nm (dash-dotted red curve), and $h=20$ nm (solid green curve). Here, we have set $\delta=0$, $a=16$ nm, and $\Omega=70$ meV. The inset represents $g^{(2)}(\tau)$ for $h=20$ nm. }
		\label{fig:9}
	\end{figure}

  In Fig.\ref{fig:9} we have plotted the normalized second-order autocorrelation function, $g^{(2)}(\tau)$, against the delay time $\tau$ for different values of the QD-MNP separation distance. We assume that the emitter is near resonance with the higher-order plasmon modes. The oscillatory behavior of $g^{(2)}(\tau)$ indicates the non-Markovianity of the system dynamics.
			As the QD-MNP separation distance increases, the amplitude and the number of oscillations decrease. In other words, as the QD-MNP separation distance decreases the backaction of the reservoir on the QD emitter and its re-excitation increases significantly. Physically, this behavior can easily be understood through the dependence of the photon-number statistics on the intensity of the LDOS. As can be seen from Figs.\ref{fig:8}(a) and \ref{fig:8}(b)), with increasing the QD-MNP separation distance from $h=4$ nm (dashed red curve) to $h=20$ nm (dash-dotted green curve) the intensity of the LDOS decreases which indicates the decreasing of the QD-MNP coupling strength. Consequently, the oscillatory behavior of $g^{(2)}(\tau)$ is decreased as well. Moreover, the QD-nanosphere hybrid system under study behaves as an ideal single-photon source ($g^{(2)}(0)=0$), irrespective of the nanosphere size.
	\section{CONCLUSIONS}
	\par 	In conclusion, we have presented a theoretical study of the photon-number statistics of a hybrid system composed of an emitter (QD) in the proximity of an MNP with non-Markovian dynamics. Our approach is based on the classical Green's function technique and a time-convolution master equation. The InAs quantum dot is taken as a three-level $\Lambda$-configuration system in which transition channels are orthogonal. One of the transition channels is along the $x$ axis and the other is along the $z$ axis. The $x$ channel is coupled to a $x$-polarized classical driving field while both channels are coupled to the elementary excitations of the MNP. The statistical properties of the emitted photons are determined through the normalized second-order correlation function which can be controlled by the geometrical as well as the physical parameters of the system, including the QD-MNP separation distance, the QD-surface plasmon modes detuning, and the Rabi frequency of the $x$-polarized driving field.
	 \par We have numerically examined the memory effects on the spontaneous decay of the QD and the dynamics of
	 the system under study. We have shown that the LDOS around the nanoshell (nanosphere) is affected dramatically by the QD-MNP separation distance as well as the materials of the core and the embedding medium. Since the MNP affects on $x$- and $z$-polarized modes differently the anisotropic Purcell effect happens. The FWHM of the LDOS, which originates from the dissipation in the system under study as well as the intensity of the LDOS, have significant effect on the second-order correlation function. Following this viewpoint, the smaller the FWHM is, the more oscillatory behavior in the photon-number statistics is. Higher intensity of LDOS leads to more oscillatory behavior of the autocorrelation function. We have found that by increasing enough the QD-MNP separation distance and/or the QD-surface plasmon mode detuning the behavior of the system enters the Markovian regime. We have also shown that in both Markovian and non-Markovian
	 regimes the emitted photons of the system exhibit the antibunching feature. Moreover, the QD-MNP hybrid system under study behaves as an ideal single-photon source $(g^{(2)}(\tau)=0)$, irrespective of the QD-MNP separation distance. To sum up, the results reveal that the presented hybrid system has the potential to be a highly controllable single-photon source.

\end{document}